\documentclass{lmcs}
\pdfoutput=1

\usepackage{lastpage}
\lmcsdoi{18}{2}{18}
\lmcsheading{}{\pageref{LastPage}}{}{}%
{Jun.~26,~2020}{Jun.~14,~2022}{}

\usepackage{diagrams}
\diagramstyle[repositionpullbacks,midshaft,scriptlabels]

\usepackage[utf8]{inputenc}
\usepackage{amsmath}
\usepackage{amssymb}
\usepackage{meq}
\usepackage{xspace}
\setlist{itemsep=.5em,topsep=.5em}

\usepackage{color}
\usepackage{colortbl}
\usepackage{rgbcolor}
\definecolor{mycolor}{gray}{0.8}
\usepackage{graphicx}
\usepackage{fancyvrb}
\usepackage{etoolbox}
\usepackage{float}

\usepackage{header}
\usepackage{framed}

\usepackage{wrapfloat}

\usepackage{booktabs}

\begin{document}
%**end of header

\title[Modular Termination for Second-Order Computation Rules]
{Modular Termination\texorpdfstring{\\}{ }for Second-Order Computation Rules\texorpdfstring{\\}{ }%
and Application to Algebraic Effect Handlers}

\author[M.~Hamana]{
Makoto Hamana\lmcsorcid{0000-0002-3064-8225}
}
\address{Faculty of Informatics, Gunma University}

\email{
hamana@gunma-u.ac.jp
}
\keywords{Theory, \lmd-calculus, algebra, modularity,
  termination, second-order algebraic theory, 
functional programming language, Haskell%
}
\subjclass{D.3.2, E.1, F.3.2}
\begin{abstract} 
We present a new modular proof method of termination for second-order
computation, and report its implementation \SOL.  The proof method is useful
for proving termination of higher-order foundational calculi. To establish
the method, we use a variation of the semantic labelling translation and
Blanqui's General Schema: a syntactic criterion of strong normalisation.
an application, we show termination of a variant of
call-by-push-value calculus with algebraic effects, an effect handler and
effect theory.  We
also show that our tool \SOLall is effective to solve higher-order
termination problems.

 \end{abstract}

\maketitle

\section{Introduction}

Computation rules such as the \beta-reduction of the \lmd-calculus
and arrangement of \fn{let}-\BR
expressions
are fundamental mechanisms of functional programming.
Computation rules for modern functional programming are necessarily higher-order and are presented as a \lmd-calculus extended with
extra rules such as rules of \fn{let}-expressions
or first-order algebraic rules like ``$0 + x \to x$''.

The termination property is one of the 
most important properties of such a calculus because it is a key to
ensuring the decidability of its properties.
A powerful termination checking method is important
in theory and in practice. 
For instance, 
Agda and Coq perform termination checking before type checking
for decidable type checking.
Haskell's type families~\cite{Mtype1,Mtype2} have
several syntactic restrictions on the form of type instances
to ensure termination,
but a more flexible and powerful termination checking method is desirable.
Although currently Haskell's rewrite rule pragma~\cite{Hrewrite} does not have
any restriction on the rules, ideally 
some conditions or termination checking are necessary,
because the compiler may go into an infinite loop without termination assurance.

In the situations listed above,
termination is ideally checked \W{modularly}. 
Several reasons for it can be given.
\begin{enumerate}[leftmargin=*]
\item A user program is usually built on various library programs.
Rather than checking termination of the union of the user and 
library program at once, 
we prove termination of library programs beforehand.
And then we just prove termination of the user program at compile time
to ensure termination of the whole program.\label{itm:lib-user}

\item For a well-known terminating calculus
(such as the typed \lmd-calculus), one often extends it by adding 
extra computation rules to enrich the computation power.
Usually, proving termination of the extended calculus directly by hand is difficult.
If merely proving termination of the extra rule part suffices to
conclude termination of the extended calculus, then
the termination proof becomes much easier. 

\item Effectful \lmd-calculus with
  {effect handlers}~\cite{CBPV,handlingEff,HandlerAction,FosterKammerEffJ}
can accommodate computational effects and ordinary values.
{An effect handler} 
interprets effect terms as actual effects.
Moreover, each effect has an \W{effect theory} required to satisfy 
\cite{PlotPowerFos}.
Such an effect theory can be regarded as optimization rules of effectful
programs.
We must ensure the termination of the combination of such optimization rules
and effectful \lmd-calculus with effect handlers.
Because the effectful \lmd-calculus with effect handler is generic,
its termination can be established beforehand, 
but an effect theory varies according to used effects in a program.
Ideally, we prove termination of effect theories merely
to ensure the termination of the whole calculus: the effectful \lmd-calculus with effect theories\footnote{In \Sec\ref{sec:eff}, we demonstrate the termination of this using 
our result of a modular termination theorem.}.

\end{enumerate}

However, in general,
strong normalisation is not modular.
There are examples which show that, given two separated terminating
computation rule sets \Ax and \Bx, the disjoint union $\Ax\uni\Bx$ is not
always terminating~\cite{ToyamaModSN}.
Therefore, the above mentioned modular termination checking
is not immediately available.

Although unrestricted modular termination checking fails,
for various {restricted} classes,
modular termination results have been obtained.
This includes the cases of first-order algebraic rules
\cite{TKB,Ohlebusch,Gramlich,moduldar-dp} and
the \lmd-calculus extended with first-order algebraic rules
\cite{Okada,cube,Breazu-Gal,PolPhD}.
There are various termination criteria for higher-order pattern rules:
\cite{IDTS00,Blanqui-TCS,NormalHORPO,Pol}.
By a higher-order pattern rule, we mean that the left-hand side
is a Miller's higher-order pattern~\cite{llam}, and
the rule is matched by higher-order matching,
not just by matching modulo \alpha-renaming. This class of rules is 
more general than ordinary higher-order functional programs,
such as Haskell's programs and rewrite rules, where the left-hand sides are first-order patterns.
This class is important to faithfully formalise foundational calculi
such as the typed \lmd-calculus and its variants 
as higher-order pattern rules.
For example, the class includes the encoding of the \beta-reduction rule
$
\fn{lam}(x.M[x]) @ N \TO M[N]
$
and the \eta-reduction rule
$
\fn{lam}(x.(L @ x)) \TO L
$
as sample higher-order pattern rules, where the metavariable $L$
cannot contain the variable $x$,
although $M$ can contain $x$
because of higher-order patterns and matching.

\subsection{The framework}

We use the framework of
 \Hi{second-order computation systems}, which 
are
computational counterparts of \Hi{second-order algebraic theories}~\cite{2ndAlg,2ndCSL}.
They 
formalise pattern rules consisting of second-order typed terms.
The limitation to 
second order 
is irrelevant to
the type structure of the object language.
Second-order abstract syntax~\cite{FPT,Fiore2nd} has been demonstrated 
capable to encode higher-order terms of any order. 
Also, a well-developed model theory exists for second-order syntax:
algebras on presheaves and \Sigma-monoids~\cite{FPT,free,CRS}.

Using this framework, we have formulated
various higher-order calculi as second-order algebraic theories
and have checked their decidability using second-order \resyss and the
tool \SOL, the second-order laboratory~\cite{SOL,pocr,jfp19,scp19}.
Staton demonstrated that second-order algebraic
theories are a useful framework
that models various important notions of programming languages such as
logic programming~\cite{SamFOS}, algebraic effects~\cite{SamInstance,Substituion-Jumps}, and
quantum computation~\cite{SamQ}. 
Our modular termination method is applicable to
algebraic theories of these applications.

\subsection{Contributions}
In this paper, we use the word \Hi{termination} to mean
strong normalisation (in short, \Hi{SN}, meaning that any computation path is finite).
We establish a modular termination proof method of the following form: 
if \Ax is SN and \Bx is SN 
with some suitable conditions, then $\RR = \Ax\uni\Bx$ is SN.
More precisely, the contributions of this paper can be
summarised as described below.

\begin{enumerate}[label={{\arabic*.}},leftmargin=2em]
\item We present a new modular proof method
of termination for second-order
computation and prove its correctness (\Sec \ref{sec:mod-th}, \Sec \ref{sec:sl}).

\item As an application of the modular
  termination proof method,
we give a termination proof of effectful calculus using a variant of
  Levy's call-by-push value (CBPV) calculus~\cite{CBPV}
called multi-adjunctive metalanguage (\MAM) \cite{FosterKammerEffJ} (\Sec 
  \ref{sec:eff})
with effect handlers and effect theory~\cite{handlingEff}.

\item As another application of the modular
  termination proof method,
we give a termination proof method for combinations 
of first-order and higher-order rules (\Sec \ref{ex:fo-split}).

\item We report the implementation SOL and experiments over a collection
  of termination problems (\Sec \ref{sec:bench}, \Sec \ref{sec:termcomp}).
\end{enumerate}

\subsec{Organisation} 
This paper is organised as follows.
We first introduce the framework of this paper, and key techniques
in \Sec \ref{sec:prelim}.
In \Sec \ref{sec:mod-th}, we prove the main theorem.
In \Sec \ref{sec:sl}, we prove a version of higher-order semantic labelling.
In \Sec \ref{sec:eff} and \Sec \ref{ex:fo-split},
we show applications of the main theorem through several examples.
In \Sec \ref{sec:related}, we discuss related work and give a summary.
The computation systems in examples of this paper 
are available at the \fn{arXiv'20} pull-down menu
of the web interface of \SOL system \url{http://solweb.mydns.jp/}\;.

 \def\th{{\mathhexbox112}\xspace}
\section{Preliminaries}\label{sec:prelim}
\subsection{Second-Order Computation Systems}\label{sec:salg}

We introduce a formal framework of 
second-order computation 
based on second-order algebraic theories~\cite{2ndCSL,2ndAlg}.
This framework has been used in~\cite{SOL}.

\Notation[def:notation]
The notation $A\uplus B$ denotes the disjoint union of two sets,
where $A\cap B = \emptyset$ is supposed.
We use the notation $\vec{A}$ for a sequence $A_1,\ccc,A_n$, 
and $|\vec{A}|$ for its length.
We abbreviate the words left-hand side as ``lhs'', right-hand side as ``rhs'',
first-order as 
``FO'' and  higher-order as ``HO''.

For a binary relation $\to$,
we write $\too$ for the reflexive transitive closure,
$\to^+$ for the transitive closure, and $\ot$ for the inverse of 
$\to$.

For a preorder $\gee$,
we write $s \gtt t$ if $s \gee t$ and $t\not \gee s$, 
and the transitive and irreflexive relation $\gtt$ is called the strict part of $\gee$.
\oNotation

\subsection{Types}
We assume that \AA is a set of \W{atomic types}
(e.g. \fn{Bool}, \fn{Nat}, etc.).
We also assume a set of 
\W{type constructors} together with arities $n\in\Nat,\, n\ge 1$.
The sets of \Hi{molecular types} \W(mol types, for short) $\TT_0$ and
\Hi{types} \TT are generated by the following rules:
\[
\infrule{Atomic}
{b \in \AA}{b\in\TT_0}
\quad
\infrule{Con}
{
  \begin{array}[h]{llll}
b_1,\ooo,b_n \in \TT_0 \\ T \;n\text{-ary type constructor}
  \end{array}
}
{T(b_1,\ooo,b_n) \in \TT_0}
\quad
\infrule{Arr}
{a_1,\ooo,a_n,b \in \TT_0 }
{a_1,\ooo,a_n \to b \in \TT}
\]

 \Remark
 Molecular types work as ``base types'' in ordinary type theories.
 But in our usage, we need ``base types'' which are
 constructed from ``more basic'' types.
 Hence we first assume atomic types as the most atomic ones, and then
 generate molecular types from them. Molecular types exactly correspond to base types in~\cite{IDTS00,SamInstance}.
 \oRemark

\subheadn{Terms}
A \Hi{signature} \Sig is
a set of function symbols of the form
$$f: (\vec{a_1}\to b_1),\ooo,(\vec{a_m}\to b_m) \to c$$
where all $a_i,b_i,c$ are mol types
(thus any function symbol is of up to second-order type).
A sequence of types may be empty in the above definition. 
The empty sequence is denoted by $\emptytype$, which may be omitted,
e.g., $b_1,\ooo,b_m \to c\, ,$ or
$\emptytype\to c$. The latter case is simply denoted by $c$.
We assume two disjoint syntactic classes of letters, called
\Hi{metavariables} (written as capital letters $\va m, \va n,\va k,\ooo$)
and  \Hi{variables} (written usually $x,y, \ooo$).
The raw syntax is given as follows.
\[
\begin{array}{lllllll}
\text{$\bullet$ \Hi{Terms} have the form}
&
t ::= x \| {x^a}.{t} \| f(t_1,\ooo,t_n).

\\
\text{$\bullet$ \Hi{Meta-terms} extend  terms to}
&
s ::= x \| {x^a}.{s} \| f(s_1,\ooo,s_n) \| {\va{m}}[s_1,\ooo,s_n].
\end{array}
\]

\noindent
These forms are respectively \W{variables},
\W{abstractions}, and \W{function terms}, and
the last form is called a \W{meta-application}.
We may write $x_1^{a_1},\ooo,x_n^{a_n}.\,t\,$ (or $\vec{x^a}. t$) for
$x_1^{a_1}.\ccc.x_n^{a_n}.\,t$, and we assume ordinary \alpha-equivalence 
for bound variables.
Hereafter, we often omit the superscript of variables $x_i^{a_i}$.
We also assume that all bound variables and free variables
are mutually disjoint 
in computation steps to avoid \alpha-renaming during computation.
If computation rules do not satisfy this property, we consider
suitable variants of the rules by renaming free/bound (meta)variables.
A metavariable context $Z$ is a sequence of (metavariable:type)-pairs,
and
a context \Gamma is a sequence of pairwise distinct
(variable:mol type)-pairs. Thus writing a context $\Gamma,\Gamma'$, 
we implicitly mean that
\Gamma and $\Gamma'$ are disjoint.
A judgment is of the form
$$
\mju{Z}\Gamma t { b}.
$$

A meta-term $t$ is called \W{well-typed} 
if $\mju Z {\Gamma} t c$ is derived by the typing rules in  Fig. \ref{fig:meta-terms}
for some $Z,\Gamma,c$.

\begin{figure}
\normalsize
\begin{meq}
\x{-1em}
\nxinfrule{Name}
{y : b\in \Gamma}
{\mju Z \Gamma y b}
\quad
\nxinfrule{Env}
{ 
\begin{array}[h]{lllll}
(\va{m}:a_1,\ooo,a_m\to b) \in Z \quad\\
 \mju Z \Gamma {t_i}{a_i} \quad 
 (1\le i \le m)
\end{array}
}
{\mju Z \Gamma {\va{m} [t_1, \ooo, t_m]} {b} }
\quad
\nxinfrule{Bind}
{ \mju Z {\Gamma,\vec{x : a}} {t} {{b}} 
}
{ \mju Z {\Gamma} {\vec{x : a}.t} {\vec a \to b}
}
\\[.5em]
\nxinfrule{Fun}
{
\begin{array}[h]{lllll}
f : (\vec{a_1}\to {b_1}),\ccc,(\vec {a_m}\to {b_m})\to c \in \Sig
\\
\mju Z {\Gamma} {\vec{x_i^{a_i}}.s_i} {{b_i}} \quad (1\le i \le m)
\end{array}
}
{ \mju{Z}{\Gamma} {f(\vec{x_1^{a_1}}.s_1, \ooo, \vec{x_m^{a_m}}.s_m)
    }{c}
}
\end{meq}
\caption{Typing rules of meta-terms}
\label{fig:meta-terms}
\end{figure}

\subhead{\Presheaves of meta-terms}In the proofs of this paper, we will 
use the structure of type and context-indexed sets.
A \Hi{\presheaf} $A$ is a family
$\set{A_b(\Gamma) \| b \in \Ty,\; \text{context }\Gamma}$ of sets
indexed by types and variable contexts.
Set operations such as $\union,\uplus,\cap$ are
extended to \presheaves by index-wise constructions, 
such as $A\union B$ by  $\set{A_b(\Gamma)\union B_b(\Gamma)\| b \in \Ty,\; \text{context }\Gamma}$.
Throughout this paper, for a \presheaf $A$,
we simply write $a \in A$ 
if there exist $b,\Gamma$ such that $a \in A_b(\Gamma)$.
The indices are usually easily inferred from context.
A \Hi{map $f : A \to B$ between \presheaves} is given by indexed functions
$\set{f_b(\Gamma) : A_b(\Gamma) \to B_b(\Gamma) \| b \in \Ty,\; \text{context }\Gamma}$.
Examples of \presheaves are 
\Hi{the \presheaves of meta-terms \MZ} and \Hi{of terms \TsigV} defined by 
\begin{meqa}
\MZ_b(\Gamma)  &\deq \set{ t \| Z \tpr \Gamma \pr t : b},  \qquad
\TsigV_b(\Gamma)  \deq \set{ t \| \tpr \Gamma \pr t : b}.
\end{meqa}
for a given signature \Sig.
We call a meta-term $t$ \W{\Sig-meta-term} if 
$t$ is constructed from \Sig and (meta)variables, i.e.,
$t\in M_\Sig Z$ for some $Z$.

\begin{figure}\normalsize
\begin{meq}
\arraycolsep = 0mm
\ninfrule{Rule}
{
\begin{array}[h]{lllll}
\mju{}{\Gamma',\vec{x_i : a_i}} {s_i} { {b_i}} \quad 
 (1\le i \le k)
\quad \theta=\mesub
\\
(\mju {\va m_1 :{(\vec {a_1}\to  {b_1})},\ooo,\va m_k :
  {(\vec {a_k}\to {b_k})}}
{}{\el \TO r } { c}) \in \RR
\end{array}
}
{\mju{} {\Gamma'} {\ext\theta(\el)  \toR \ext\theta(r)} { c}}
\\[.7em]
\ninfrule{Fun}
{
\begin{array}[h]{lllll}
f : (\vec{a_1}\to    {b_1}),\ccc,(\vec {a_k}\to     {b_k})\to  c \in \Sig
\\
\mju{} {\Gamma,\vec{x_i : a_i}} {t_i\toR t'_i} { {b_i}} \quad 
 (\text{some single }i\text{ s.t. }1\le i \le k)
\end{array}
}
{\mju{} {\Gamma} 
{f(\vec{x_1^{a_1}}.t_1,\ooo,\vec{x_i^{a_i}}.t_i\ooo,\vec{x_k^{a_k}}.t_k) \toR
 f(\vec{x_1^{a_1}}.t_1,\ooo,\vec{x_i^{a_i}}.t'_i\ooo,\vec{x_k^{a_k}}.t_k) } {     c}}
\end{meq}
\caption{Second-order computation (one-step)}
\label{fig:so-com}
\end{figure}

The notation $t\subi{x_1\mapsto s_1,\ooo,x_n\mapsto s_n}\; $
denotes ordinary capture avoiding substitution that replaces
the variables with terms $s_1,\ooo,s_n$.
\DefTitled[def:msubst]{Substitution of terms for metavariables}
Let $n_i=|\vec{c_i}|$ and 
$\vec{c_i} = c_i^1,\ooo,c_i^{n_i}$.
Suppose 
\begin{meqa}
\mju {}{\Gamma',\, x_i^1\:c_i^1,\ooo,x_i^{n_i}\:c_i^{n_i}
}{&s_i}{    {b_i}}\qquad (1\le i\le k),
\\
\mju{\va m_1 :{\vec{c_1}\to     {b_1}},\ooo,
   \va m_k :{\vec{c_k}\to     {b_k}}}{\Gamma} {&e}      c
\end{meqa}
For an assignment
$\th=\set{\va m_1 \mapsto \vec{x_1}.s_1,\ooo,\va m_k \mapsto \vec{x_k}.s_k}
: Z \to \TsigV$,
\\
the map  $\ext\th : \MZ  \to \TsigV$ is \W{a substitution for metavariables}
defined by
\[
\arraycolsep = 1mm
\begin{array}[h]{rclllllllllllllllllll}
\thAP{x} &\deq& x \\
\va \thAP{M_i[t_1,\ooo,t_{n_i}]}  &\deq& 
  s_i\;  \{ {x_i^1} \mapsto \thAP{t_1},\ooo,
{x_i^{n_i}} \mapsto \thAP{t_{n_i}} \} \\
\thAP{f(\vec{y_1}.t_1,\ooo,\vec{y_k}.t_k)}  &\deq& 
f(\vec{y_1}.\thAP{t_1},\ooo,\vec{y_k}.\thAP{t_k})
\end{array}
\]
\oDef

\Lemma[th:tsub]{\tsub}{
Under the situation of the above definition, 
the substituted term is well-typed as
 $\;\mju  {} {\Gamma,\Gamma'} {\thAP{e}}  c.$
\oLemma}
\Proof This is proved by straightforward induction on the typing derivations
\cite{2ndCSL}.
\QED

\Remark\label{rem:metaterms}
The map  $\ext\th : \MZ  \to \TsigV$
is the substitution operation of terms for metavariables, and
is a homomorphic extension of $\th : Z \to \TsigV$, depicted as
the diagram of Figure~\ref{fig:extTh}, 
\begin{figure}[ht]
    \begin{diagram}[2em]
      Z &\rTo^{\eta_Z}& \MZ\\
      &\rdTo_\theta &\dTo_{\theta^\sp}\\
      &             & \TsigV
    \end{diagram}
  \caption{The map $\ext\th$}
  \label{fig:extTh}
\end{figure}
where
$\eta_Z$ embeds metavariables to meta-terms.
It shows that $\ext\th$ is a 
unique \Sig-monoid morphism that extends \th
\cite[Def. 11, $\th^*$]{free}~\cite[III]{PolyTh}.
The syntactic structure of meta-terms and substitution
for abstract syntax with variable binding
was first introduced  by  Aczel~\cite{Aczel}.  This  formal
language  allowed  him  to  consider  a  general  framework  of
rewrite rules for calculi with variable binding. 
The structure of meta-terms has clean algebraic properties.
The \presheaf \TsigV forms an initial \Sig-monoid~\cite{FPT} and
\MZ forms a free \Sig-monoid over $Z$~\cite{free,Fiore2nd}. 
These algebraic characterisations have been applied to 
the complete algebraic charactersisation of 
termination of second-order rewrite systems~\cite{CRS} 
and higher-order semantic labelling~\cite{HSL}.
The polymorphic and precise algebraic characterisation
of abstract syntax with binding and substitution were given in~\cite{PolyTh}.
\oRemark

\subhead{Computation rules}
First we need
the notion of Miller's {second-order pattern}~\cite{llam}. 
A \Hi{second-order pattern} is
a meta-term in which every occurrence of meta-application
is of the form
$
M[x_1,\ooo,x_n],
$
where $x_1,\ooo,x_n$ are distinct bound variables.

For meta-terms $\mju{Z}{}{ \ell }{b}$ and 
$\mju{Z}{}{ r }{ b}$ using a signature \Sig,
a \W{computation rule} is of the form
$$
\mju{Z}{}{\ell \TO r }{b}
$$
satisfying:
\begin{enumerate}[leftmargin=1.5em]
\item $\ell$ is a function term and a second-order pattern.

\label{itm:lhs-dsp}

\item all metavariables in $r$ appear in \el.
\end{enumerate}
Note that $\ell$ and $r$ are meta-terms without free variables,
but may have free \Hi{meta}variables.
A \Hi{computation system} (CS) is a pair $(\Sig,\;\RR)$ of a signature
and a set \RR of computation rules consisting of \Sig-meta-terms.
We write $s \toR t$ to be one-step
computation using \RR obtained 
by the inference system given in Fig. \ref{fig:so-com}. 
We may omit some contexts and type information of a judgment,
and simply write it as $Z\tpr {\ell \TO r} : b$, ${\ell \toR r}$, or ${\ell \TO r}$
if they are clear from the context.
From the viewpoint of pattern matching, (Rule) means that
a computation system uses the decidable second-order pattern matching 
\cite{llam} for one-step computation (cf.~\cite[Sec.6.1]{SOL})
not just syntactic matching.
We regard $\toR$ to be a binary relation on terms.

A function symbol $f\in\Sig$ is called \Hi{defined}
if it occurs at the root of the lhs of a rule in \RR.
Other function symbols in \Sig are called \Hi{constructors}.

\Example[ex:SigStl]
The simply-typed \lmd-terms on the set $\Typ$ of
simple types generated by a set of base types $\BTy$
are modeled in our setting as follows.
Let $\AA = \BTy$.
We suppose type constructors \LL, \Arr.
The set of \Typ of all simple types for the \lmd-calculus
is the least set satisfying
\[
\Typ = \BTy \union \set{\Arr(a,b) \| a,b \in \Typ}.
\]
We use the mol type $\LL(a)$
for encoding \lmd-terms of type $a\in\Typ$.
The \lmd-terms are given by a signature 
\[
\Sig_{\mathrm{stl}} = \left\{
  \begin{array}{lll}
  \fn{lam}_{a,b} &: (\LL(a) \to \LL(b)) \to \LL(\Arr(a,b))\\
  \fn{app}_{a,b} &: \LL(\Arr(a,b)),\LL(a)\to \LL(b)
  \end{array}
\| a, b \in \Typ \right\}
\]
The \beta-reduction law is presented as 
\begin{meqa}
\s{(beta)}
 &{\;\;M\:\LL(a)\to \LL(b),\; N\:\LL(a)\;} \tpr\pr
 \fn{app}_{a,b}(\fn{lam}_{a,b}(x^{{\LL(a)}}.\, M[x] ),\,N)
\,\; \TO  \;\, M[N] : {\LL(b)}
\end{meqa}
Note that $\LL(\Arr(a,b))$ is a mol type, but
$a\to b$ is not a mol type. 

We use the following notational convention throughout the paper.
We will present a signature
by omitting mol type subscripts $a,b$ 
(see also more detailed account~\cite{pocr}).
For example, simply writing function symbols
\fn{lam} and \fn{app}, we mean 
$\fn{lam}_{a,b}$ and $\fn{app}_{a,b}$ in $\Sig_{\mathrm{stl}}$
having appropriate mol type subscripts $a,b$.
\oExample

\subsection{The General Schema}\label{sec:GS}
\TGS is a criterion for proving strong normalisation of higher-order rules
developed by Blanqui, Jouannaud and Okada~\cite{IDTS} and refined by Blanqui
\cite{IDTS00,Blanqui-TCS}.
We summarise the definitions and properties of \GS
in~\cite{IDTS00,Blanqui-TCS}.
\TGS has succeeded in proving SN of various rewrite rules such as 
G\"{o}del's System T.
The basic idea of \GS is to check whether the arguments of 
recursive calls in the right-hand side of a rewrite rule 
are ``smaller'' than the left-hand sides' ones. It is similar 
to Coquand's notion of 
``structurally smaller''~\cite{CoqPat}, but more relaxed and extended.
This section reviews the definitions and the property of \GS criterion~\cite{IDTS00,Blanqui-TCS}.

We give a summary of 
the General Schema (\GS) criterion~\cite{IDTS00,Blanqui-TCS}
of the second-order case.
Suppose that
\begin{itemize}\item a well-founded preorder $\le_\BB$ on the set \BB of types 
and
\item a well-founded preorder $\le_\Sig$ on a signature \Sig
\end{itemize}
are given.
Let $\lt_\BB$ be the strict part of $\le_\BB$,
and $=_\BB \;\deq\; \le_\BB \cap \ge_\BB$ the associated equivalence relation.
Similarly for $\le_\Sig$.

The \W{stable sub-meta-term ordering} $\tleS$ is defined by
$s \tleS t$ if $s$ is a sub-meta-term of $t$ and all the free variables in $s$
appear in $t$.

\Def[def:acc]
A metavariable $\va m$ is \W{accessible in a meta-term $t$} if
there are distinct bound variables $\vec x$ such that ${\va{m}}[{\vec x}]
\in \Acc(t)$, where $\Acc(t)$ is the least set satisfying
the following clauses:
\begin{enumerate}[leftmargin=*,label={{(a\arabic*)}}]
\item $t \in \Acc(t)$.
\item If ${x}.{u} \in \Acc(t)$ then $u \in \Acc(t)$.

\item Let $f:\tau_1,\ooo,\tau_n\to b$ and $f(u_1,\ooo,u_n) \in \Acc(t)$,\\
where $\tau_i = \vec{a_i} \to c_i$ for each $i =1,\ooo,n$
($|\vec{a_i}|$ is possibly $0$, and then $\tau_i=c_i$).\\
If $(\text{for all }a \in \set{\vec{a_i}},\;
a \lt_\BB b) \;\,\&\;\, c_i \le_\BB b$,\;\,
then $u_i\in \Acc(t)$.
\label{itm:acc3}
\end{enumerate}
\oDef
Note that 
\ref{itm:acc3} also says that
any of $\vec{a_i}$
must not be equivalent (w.r.t. $=_\BB$) to $b$.

\Def[def:ccl]
Given $f \in \Sig$,
the \W{computable closure} $\CCl_f(\vec t)$ of a meta-term $f(\vec t)$ is
the least set \CCl satisfying the following clauses.
All the meta-terms 
and abstractions
below are assumed to be well-typed.
\begin{enumerate}[leftmargin=*,label={\quad{\bf \arabic*.}}]
\item \textbf{(meta $\va m$)\;\;} If $\va{m} : \tau_1,\ooo,\tau_p \to b$ is accessible in 
some of $\vec t$, and $\vec u \in \CCl$,
then $\va m [\vec u]\in \CCl$.

\item For any variable $x$, $x\in \CCl$.

\item If $u\in \CCl$ then $x.u\in \CCl$.

\item \textbf{(fun $f \gt_\Sig g$)\;\;} If $f \gt_\Sig g$ and $\vec w\in \CCl$, then $g(\vec w)\in \CCl$.

\item \textbf{(fun $f =_{\Sig} g$)\;} If $\vec u\in \CCl$ such that
$\vec t \gtSmul \vec u$, then $g(\vec u)\in \CCl$, where $\gtSmul$ is
the lexicographic extension of the stable sub-meta-term ordering $\tgtS$.
\end{enumerate}
\oDef

The labels \textbf{(meta $\va m$)} etc. are used for references
in a termination proof using \GS.

\begin{thmC}[\cite{IDTS00,Blanqui-TCS}]\label{th:GS}\rm
Let  $(\Sig,\RR)$ be a \resys. Suppose that $\le_\BB$ and $\le_\Sig$ are
well-founded.
If for all $f(\vec t)\TO r \in \RR$,
$\CCl_f(\vec t) \ni r $,
then
$\RR$ 
is strongly normalising.
\end{thmC}

\Example[ex:rec]
We consider a \resys~\s{recursor} of a recursor on natural numbers. 
The signature $\Sig_{\mathrm{rec}}$ 
\cite{Blanqui-TCS} is given by
\begin{solT} 
 zero : L(Nat),   succ : L(Nat) $\to$ L(Nat)
 rec$_a$ : L(Nat),L($a$),(L(Nat),L($a$) $\to$ L($a$)) $\to$ L($a$)
\end{solT}
where $a \in \Typ$, $\s{Nat} \in \BTy$,
and \BTy and \Typ are the ones used in Example \ref{ex:SigStl}.
We take preorders $\le_\BB, \le_\Sig$ to be the identities.
The rules are
\begin{solT} 
 (recZ) rec(zero, U,x.y.V[x,y])   $\TO$ U
 (recS) rec(succ(X),U,x.y.V[x,y]) $\TO$ V[X, rec(X,U,x.y.V[x,y])]
\end{solT} 
We check $\CCl_\s{rec}(\s{succ(X),U,x.y.V[x,y]})\ni 
\s{V[X, rec(X,U,x.y.V[x,y])]}$.
\begin{itemize}[leftmargin=*]
\item \textbf{(meta \s{V})} is applicable. We check 
that \s V is accessible and $\s X \in \CCl$.
  Now
$\s{V[x,y]} \in\Acc(\s{x.y.V[x,y]})$
holds by (a1)(a2). $\s X\in\CCl$ holds because \s X is accessible
in \s{succ(X)}.

\item To check $\s{rec(X,U,x.y.V[x,y])}\in\CCl$, \textbf{(fun \s{rec}=\s{rec})} is applicable.\\
Since \s{succ(X) $\tgtS$ X} and \s{U,x.y.V[x,y]$\in\CCl$}, 
which is easily checked,
we are done.
\end{itemize}
We also check $\CCl_\s{rec}\s{(zero,U,x.y.V[x,y])} \ni \s{U}$ by \textbf{(meta \s{U})}.
Hence the computation system is SN.
\oExample

 \subsection{Higher-Order Semantic Labelling}\label{sec:hsl}

As we have seen in Example \ref{ex:rec},
\GS checks syntactical decreasing (\s{succ(X) $\tgtS$ X}) of an argument
in each recursive call.
But sometimes recursion happens with syntactically larger but semantically 
smaller arguments. For example, consider the following \resys of
computing the prefix sum of a list

\medskip
\begin{solT}
  map(y.F[y],nil)        $\TO$ nil
  map(y.F[y],cons(X,XS)) $\TO$ cons(F[X],map(y.F[y],XS))
  ps(nil)                $\TO$ nil 
  ps(cons(X,XS))         $\TO$ cons(X,ps(map(y.X+y,XS)))
\end{solT}
\medskip
In the final rule,
\s{ps} in rhs is called with a shorter list than \s{cons(X,XS)} in lhs,
but syntactically, \s{map(y.X+y,XS)} is not a sub-meta-term of \s{cons(X,XS)}.
Therefore, the rule does not satisfy \GS.
The \W{higher-order semantic labelling method}~\cite{HSL} 
extending~\cite{Zantema} solves this problem,
which is founded on the presheaf models of second-order syntax
\cite{FPT,Fiore2nd,free} and computation~\cite{CRS}.
We will use a version of the method in the proof of the main theorem, hence 
we sketch the idea of it.
The following notion is needed:
a \Hi{quasi-model} $(A,\ge)$ 
of a second-order \resys \RR is a second-order algebra equipped with a preorder 
(i.e. a weakly monotone \Sig-monoid~\cite{CRS,HSL}) in which, for every 
rule $Z \tpr \pr \el \TO r$ of \RR, 
$
\den{\el}\phi \ge \den{r}\phi
$
holds for every assignment $\phi : Z\to A$, 
where $\den{-}\phi$ is an interpretation of meta-terms using \phi.
It is called \W{well-founded} (or SN) if $\ge$ is well-founded.
If one finds a quasi-model for a given \RR,
then one attaches semantic elements in the quasi-model to the function
symbols in the rules of \RR. 
For the case of the rules of the prefix sum,
one chooses a quasi-model of natural numbers with the usual order
(where $\s{ps}$ is interpreted as counting the length of the argument list).
Applying the higher-order semantic labelling method, one obtains labelled rules
\cite{SL-CRS}:
\medskip
\begin{solLmath}
ps$_{n+1}$(cons(X,XS)) $\TO$ cons(X,ps$_{n}$(map(y.X+y,XS)))
\end{solLmath}
\medskip
for all $n\in\Nat$, where $n$ is a label.
The labelling, in principle, does not change the computational
behavior, but it is effective to prove SN because 
the call relation $\s{ps}_{n+1} \gt \s{ps}_{n}$ is well-founded,
the rule can satisfy \GS. 

The main theorem~\cite[Thm.3.7]{HSL}
of HO semantic labelling 
ensures that if the labelled second-order computation system combined with
additional \W{decreasing rules}
\medskip
\begin{solLmath}
ps$_{n+1}$(X) $\TO$ ps$_{n}$(X)
\end{solLmath}
\medskip
(which expresses compatibility of the computation relation 
$\toR$ with the order $n+1 \gt n$ of the quasi-model)
is proved to be SN, then the original system is SN.

 \renewcommand{\tltS}{{\tlt}} \renewcommand{\tgeS}{{\tge}} \renewcommand{\tgtS}{{\tgt}} \renewcommand{\tleS}{{\tlt}} 

\section{A Modular Termination Theorem for Second-Order Computation}\label{sec:mod-th}

In this section, we prove the main theorem of this paper.
In the rest of this paper, we assume the following.
{$\Fun(-)$ denotes the set of all function symbols in its argument.}

\Ass[def:setting]
    Let
    $(\SigA\uni\OH,\Ax)$ and $(\SigA\uni\SigB\uni\OH,\QQ)$ be 
    computation systems satisfying:
    \begin{enumerate}\item \RRRsig is the set of defined function symbols of \RRR.
          \item \QQsig is the set of defined function symbols of \QQ.
          \item \OH is the signature for constructors of \RR, where
$\RR \deq \RRR \uplus \QQ$.
\item \RR is finitely branching. \label{ass:fb}

\item Both sides of each rule in \RR satisfy the \SigA-layer condition.
\label{ass:fA}

\end{enumerate}
\oAss

\Def
We say that \Hi{a meta-term $u$ satisfies the \SigA-layer condition}
if 
for every $f(\vec{\vec x.t}) \;\tlt\; u$ with $f\in\SigA$,
$\Fun(f(\vec{\vec x.t})) \subseteq \SigA\uni\OH$ and 
$\vec t$ are second-order patterns.
\oDef

Note that it does not mean
$u$ is a pattern. The condition merely requires \SigA-headed sub-meta-terms
of $t$ to be patterns,
and $u$ may contain \SigB-symbols whose
arguments need not to be patterns.

Note also that every sub-meta-term of a second-order pattern is again a second-order 
pattern. A bound variable (e.g. $x$)
might become free in a sub-meta-term (e.g. $M[x]$)
of a second-order pattern (e.g. $f(x.M[x])$),
but $x$ is ``originally'' a bound variable, therefore 
it is regarded as a bound variable
in the condition of ``distinct bound variables''.
So, $M[x]$ is a second-order pattern.

\begin{figure}
\begin{center}
\includegraphics[scale=.58]{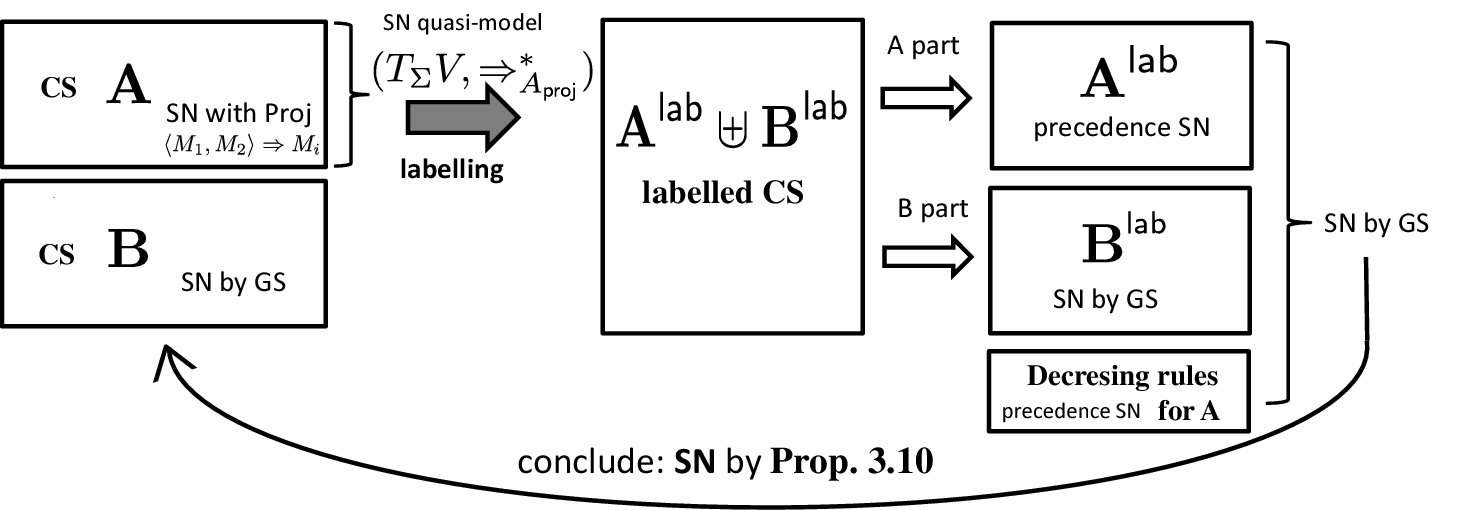}
\end{center}
\caption{Proof strategy of Thm. \ref{th:modI}
}
\label{fig:mod-cs}\end{figure}

\Remark[rem:ass]
As a consequence of this assumption,
\begin{enumerate}[label={(\arabic*)}]
\item the rhs of each rule in \Ax is a second-order
pattern,  and
\item \SigA-function symbols may appear in \Bx. \label{item:hi}
\end{enumerate}
This kind of combination of two systems 
is called \W{shared constructors~\cite{shared,Gramlich}} (i.e. constructors \OH
are shared in \Ax and \Bx), and
\W{hierarchical combination~\cite{Hie}} (because 
the \resys $(\SigA\uni\SigB\uni\OH,\QQ)$ can involve
function symbols defined in \Ax,
but $(\SigA\uni\OH,\Ax)$ cannot involve \SigB-symbols).
The technical assumption \ref{ass:fA} is needed to establish 
a variation of the HO
semantic labelling method. Technically, it is essential to make Lemma~\ref{flem:thirteen} and~\ref{fth:plab-mon} hold.
In practice, this assumption admits
effect handler examples (\Sec \ref{sec:eff}).
\oRemark

\subsection{Proof method}\label{sec:pr-method}
The modular termination theorem we want to establish is of the form: 
under Assumption \ref{def:setting},
if \Ax is SN and \Bx is SN (with some more conditions on both SN), then $\RR = \Ax\uni\Bx$ is SN.
Since in general, the assumption of SN (of each of \Ax and \Bx) is not necessarily  established by \GS, 
an idea to prove SN of \RR is
\begin{enumerate}
\item to use the precedence termination (of \Ax or \Bx, or both) by HO labelling~\cite[Thm.~5.6]{HSL},
and 
\item then to use \GS to prove SN of the labelled \RR.
\end{enumerate}
\W{Precedence termination} is a notion of termination:
if \RR is accessible and
a well-founded relation $\gtt$ exists such that
for every rule $f(\vec t)\TO r \in \RR$, every function symbol $g$ in $r$,
$f\gtt g$ holds, then \RR is called {precedence terminating}, which 
implies SN.

But to apply HO semantic labelling,
one must first seek a \W{well-founded quasi-model} of 
the whole system $\RR$.
A natural candidate of quasi-model for $\RR$
is $(\TsigV, \TO_{\Ax\uplus\Bx}^*)$,
i.e., all terms with the many-step computation relation using \Ax and \Bx,
but proving that it is \W{well-founded} is difficult.
In fact, it is nothing but the termination property (i.e. $\RR$
is SN) we are seeking to prove.
This means that the original HO semantic labelling is not quite appropriate.
To overcome the difficulty, in this paper, we use 
a variation of the semantic labelling (\Sec\ref{sec:sl}).
Instead of requiring a quasi-model of the whole rules,
we now require only a quasi-model of \Ax.
This gives a well-founded quasi-model $(\TsigV, \tooAbot)$ of \Ax
(see Fig. \ref{fig:mod-cs}),
where ${\Abot}$ is an extension of \Ax with 
projections of pairs $<M_1,M_2> \TO M_i \; (i=1,2)$ (Def. \ref{def:aux}).

Attaching a \SigA-headed term to each \SigA-function symbol
as labels in \RR, we construct a labelled \resys $\labed\Ax\uni\labed\Bx$.
Then we try to prove SN of it by \GS. To do so,
we presuppose that \Bx should be SN by \GS.
Then the $\labed\Ax$-part is precedence terminating
by labelling, and the $\labed\Bx$-part is SN by GS by assumption.

Finally, 
a variation of the HO semantic labelling theorem (Prop. \ref{th:pred-lab}) ensures that
SN of the labelled \resys $\labed\Ax\uni\labed\Bx$ (with additional decreasing rules) implies
SN of the original $\Ax\uni\Bx$. 
This establishes a modular termination theorem (Thm. \ref{th:modI}).

\subsection{Labelled rules}

\DefTitled[def:aux]{Projection rules}
We extend the \resys \Ax with pairs and the ``projection rules'' on 
the pairs.
For every mol type $b\in\BB$,
we use pairing
$<-,->_b $  and a bottom element $\bot_b$.
\begin{itemize}
 \item $\Sig_\Proj \deq \set{\bot_b : b \| b \in\BB}\uni
\set{<-,->_b : b,b\to b \| b  \in\BB}$

\item $\Proj \deq 
    \set{M_1,M_2 :b \tpr \pr <M_1,M_2>_b \TO M_1 :  b,$\\
\phantom{K}\qquad\quad
$M_1,M_2 :b \tpr \pr <M_1,M_2>_b \TO M_2 :  b 
      \| b\in \BB}$.
\end{itemize}
\oDefTitled

We define
$$
\Sig \deq \RRRsig\uplus\QQsig\uplus\OH\uplus\Sig_\Proj,\;
    \RX \deq \RRR \uplus \fn{Proj}.
$$
We construct a labelled \RR.
We label function symbols in \SigA,
and do not label other function symbols,
where labels are taken from $\TsigV$. We define the labelled signature by
$$
\LSig = \set{f_u \| f\in\SigA,\; u \in \TsigV} \union 
\SigB\uni\OH\uni\Sig_\Proj.
$$
Labelled terms are constructed by the typing rules 
in Fig. \ref{fig:meta-terms} using \LSig,
instead of \Sig.
The \presheaf $\MlabZ$ of labelled meta-terms is defined by collecting 
all labelled meta-terms.

Next we define a labelling map $\lab\phi$
that attachs labels to plain meta-terms using a function $\aext\phi$
to calculate the labels.

Let $\phi : Z \to \TsigV$ be an assignment.
The \W{term labelling} $\lab\phi$ is defined using 
$\aext\phi$, i.e.,
$\nmlab : \MZ \rTo \MlabZ$ 
is a map defined by \[
\arraycolsep = .5mm
\begin{array}[h]{llllll}
    \nmlab(x) &= x\\
    \nmlab(M[t_1,\ooo,t_n])
    &= M[\nmlab(t_1),\ooo,\nmlab(t_n)]
\\
    \nmlab( f(\vec{x_1}. t_1,\ooo, \vec{x_n}. t_n ))
&= f_{ \aext\phi{(f(\vec{x_1}.t_1,\ooo,\vec{x_n}.t_n))}}
    (\vec{x_1}.\nmlab(t_1),\ooo,\vec{x_n}.\nmlab(t_n) )
&&\text{if $f\in \Sig_\RRR$}
\\
    \nmlab( f(\vec{x_1}. t_1,\ooo, \vec{x_n}. t_n ))
&= f
    (\vec{x_1}.\nmlab(t_1),\ooo,\vec{x_n}.\nmlab(t_n) )
&&\text{if $f\not\in \Sig_\RRR$}
\end{array}
\]

\noindent
We define labelled rules
\begin{meqa}
    \labed\RR &\deq \set{Z \tpr \lab\phi(l) \TO  \lab\phi(r) :b
      \; \|  Z \tpr\; l\TO r :b \in\RR,
      \text{ assignment }\phi : Z \rTo \TsigV}
\end{meqa}
We only attach labels to \SigA-symbols.
\label{sec:trace-labelling}

\Def[def:decl]
The set of \W{decreasing rules} $\Decl$ consists of rules 
\[
f_v(\vec{x_1}.{{M_{1}}[\vec{x_1}]},\;\ooo,\;
  \vec{x_n}.{{M_{n}}[\vec{x_n}]})
\;\TO\;
f_w(\vec{x_1}.{{M_{1}}[\vec{x_1}]},\;\ooo,\;
  \vec{x_n}.{{M_{n}}[\vec{x_n}]})
\]
for all $f\in\SigA,\, v,w\in\TsigV$ with all $v \;\AprojSUB\; w$.
Here $\tltS$ is the strict subterm relation.
\oDef

\subsection{Traces and labelled systems}

\Def
We define a list-generating function $\tuple_b$ taking a finite set $\set{t_1,\ooo,t_n}$
of terms of type $b$ and returning
a tuple using the pairing as
$$
\tuple_b(\emptyset) \deq \bot_b;\quad
\tuple_b(S) \deq <\,t,\;\tuple_b(S - \set{t})\,>_b
$$
where we pick a term $t\in S$ by some fixed order on terms
(such as lexicographic order of alphabetical order on symbols).
This returns a tuple as
$\tuple_b(\set{t_1,\ooo,t_n}) = <t_1,<t_2,\ooo,<t_{n},\bot>>>.$
If $S$ is an infinite set, $\tuple_b(S)$ is undefined.
Hereafter, we will omit writing the 
type subscripts of $\bot,<-,->,\tuple$,
which can be recovered from context.
\oDef

Applying \Proj rules, we have
\begin{meqa}
\tuple(\set{t_1,\ooo,t_n}) &= <t_1,<t_2,\ooo,<t_{n},\bot>>> \tooRX t_i,\\
\tuple(\set{t_1,\ooo,t_n}) &= <t_1,<t_2,\ooo,<t_{n},\bot>>> \tooRX \bot.
\end{meqa}
The reason why we need the pairs
is to make a term constructed by a trace of reduction,
i.e., by collecting terms appearing in a reduction.
Projection rules are used to pick a term from such a trace.

We use another function $\trace(t)$ to calculate a label,
which collects all the traces of 
reductions by \RR starting from a \SigB-headed term. 
The \W{trace map}
$\trace :\TsigV \rTo \TsigV$
is defined by \[
\arraycolsep = .5mm
\begin{array}[h]{lllllll}
    &\trace(x) &=  x \\
&\trace( f(\vec{\vec{x}.t}) )&=  
    f(\vec{x_1}.\trace(t_1),\ooo,\vec{x_n}.\trace(t_n))
\quad&\text{if $f\not\in \SigB$}
\\
    &\trace( f(\vec{\vec{x}.t}) )&= 
    \tuple(\set{\,\trace(u) \| f(\vec{\vec{x}.t}) \TRACEtoR u \trCond
\,}
)
    &\text{if $f\in \SigB$}
\end{array}
\]

\Remark
The Assumption \ref{def:setting} \ref{ass:fb} that \RR is finitely
branching is necessary
to ensure a finite term constructed by the second clause.
If $f(\vec{\vec{x}.t})$ with $f\in \SigB$ is a normal form, then
$\trace( f(\vec{\vec{x}.t}) ) = \bot$.

The notion of $\trace$ and the method of
transforming a minimal non-terminating \RR-reduction sequence 
to a reduction sequence using the projection rules
has been used in proving modularity of termination~\cite[Def. 3]{Gramlich}\cite{Ohlebusch}.
\oRemark
The \W{trace labelling} $\labtr : \TsigV \rTo \Tlab$ is defined by
$\labtr \deq \lab{\trace \o \psi}$
with the unique map $\psi:\emptyset \to \TsigV$.
It satisfies:
\[
\arraycolsep = .5mm
\begin{array}[h]{llllll}
    \labtr(x) &= x\\
    \labtr( f(\vec{x_1}. t_1,\ooo, \vec{x_n}. t_n ))
&= 
    f_{ f(\vec{x_1}.\trace{(t_1)},\ooo,\vec{x_n}.\trace{(t_n)}) }
    (\vec{x_1}.\labtr(t_1),\ooo,\vec{x_n}.\labtr(t_n) )
&&\text{if $f\in \Sig_\RRR$}
    \\
    \labtr( f(\vec{x_1}. t_1,\ooo, \vec{x_n}. t_n ))
    &= f
    (\vec{x_1}.\labtr(t_1),\ooo,\vec{x_n}.\labtr(t_n) )
&&\text{if $f\not\in \Sig_\RRR$}
\end{array}
\]

\DefTitled[def:lmsubst]{Labelled substitution for metavariables}
We define
$
\sap t \rho = \labtrUN{t\rho}.
$
Given an {assignment} for labelled meta-terms 
$\th : Z \rTo \Tlab$, {labelled substitution for metavariables} 
$\lext{\th} : \MlabZ \rTo \Tlab$
is defined by \[
\arraycolsep = 1mm
\begin{array}[h]{rclllllllllllllllllll}
    \lext\th(x)  &=& x \\
    \lext\th( f(\vec{y_1}.t_1,\ooo,\vec{y_m}.t_m) ) &=& 
    f(\vec{y_1}.\lext\th(t_1),\ooo,\vec{y_m}.\lext\th(t_m)  )
\quad \text{ if }f \in\labed\Sig
    \\
\lext\th (\va m[\vec t]) )   &=& 
\sAP{\vec x}u{\vec t}
\quad\text{for }\th:M \mapsto \vec{x}.u.
\end{array}
\]
Here, the term $\forget t$ is obtained by
deleting all labels in a labelled term $t$.
\oDef

\begin{figure}
\begin{meq}
\arraycolsep = 0mm
\ninfrule{Rule}
{
\begin{array}[h]{lllll}
\mju{}{\Gamma',\vec{x_i : a_i}} {s_i} { {b_i}} \quad 
 (1\le i \le k)
\quad
\th = \mesub
\\
(\mju {\va m_1 :{(\vec {a_1}\to  {b_1})},\ooo,\va m_k :
  {(\vec {a_k}\to {b_k})}}
\Gamma{\el \TO r } { c}) \in \labed\RR
\end{array}
}
{\jud {\Gamma,\Gamma'} {\lext\th(\el)  \toLabC \lext\th(r) } { c}}
\\[.7em]
\ninfrule{Abs}
{
\jud {\Gamma,\vec{x : a}} {t\toLabC t'} { {b}} 
}
{\jud {\Gamma} 
 {\vec{x^{a}}.t
\toLabC
 \vec{x^{a}}.t' } {\vec a \to c}
}
\\[.7em]
\ninfrule{Fun}
{
\begin{array}[h]{lllll}
f : (\vec{a_1}\to    {b_1}),\ccc,(\vec {a_k}\to     {b_k})\to  c \in \labed\Sig
\qquad \text{($f$ may be labelled)}\\
\jud {\Gamma} {\vec{x_i^{a_i}}.t_i\toLabC \vec{x_i^{a_i}}.t'_i} { \vec{a_i}\to{b_i}} \quad 
 (\text{some single }i\text{ s.t. }1\le i \le k)
\end{array}
}
{\jud {\Gamma} 
{f(\vec{x_1^{a_1}}.t_1,\ooo,\vec{x_i^{a_i}}.t_i\ooo,\vec{x_k^{a_k}}.t_k) 
\toLabC
 f(\vec{x_1^{a_1}}.t_1,\ooo,\vec{x_i^{a_i}}.t'_i\ooo,\vec{x_k^{a_k}}.t_k) } {     c}}
\end{meq}
                                
\caption{Labelled reduction}
\label{fig:so-com-lab}
\end{figure}

 The labelled reduction on labelled terms 
$s \tolR t$
is defined by the inference system given in Fig. \ref{fig:so-com-lab}.

We can also apply 
the GS defined 
in \Sec \ref{sec:GS}
to prove SN of labelled systems. Appendix \ref{sec:lgs} 
proves the validity.

\Prop[th:lGS]{\lGS}{
Let  $(\labed\Sig,\labed\RR)$ be a labelled \resys. Suppose that $\le_\BB$ and $\le_{\labed\Sig}$ are
well-founded.
If 
for all $f(\vec t)\TO r \in \labed\RR$,
$\CCl_f(\vec t) \ni r $,
then
$\labed\RR$ with $\tolR$
is strongly normalising.
Note that $f$ may be labelled.
\oProp}

The following is 
a variation of the higher-order semantic labelling explained in \Sec
\ref{sec:hsl} and proved
in Appendix \ref{sec:sl}.

\Prop[th:pred-lab]{\PredLab}{
If $\labed\RR\union\Decl$ is SN, then \RR is SN.
\oProp}

\subsection{Proving a modular termination theorem}

Then we prove the main theorem.
We say that \W{\Ax is accessible} if
for each $f(\vec {\vec x.t}) \TO r \in \Ax$,
every metavariable occurring in $r$ is accessible
in some of $\vec t$.

\ThTitled[th:modI]{Modular Termination}
Let $(\SigA\uni\OH,\Ax)$ and $({\SigA}\uni\SigB\uni\OH,\QQ)$ be
computation systems satisfying Assumption \ref{def:setting} and the following.
\begin{enumerate}[label={({\roman*})}]

\item \Ax is accessible.
      \label{itm:as-acc}
      \label{itm:I-rhs-pat}
\item $(\RRRsig\uplus\QQsig\uplus\OH\uplus\Sig_\Proj,\Aproj)$ is SN (not necessarily by \GS). \label{itm:I-A-SN}
\item $({\SigA}\uni\QQsig\uplus\OH,\QQ)$ is SN by \PGS. \label{itm:I-B-SN}
\end{enumerate}
Then $(\RRRsig\uplus\QQsig\uplus\OH,\, \Ax\uplus\Bx)$ is SN.
\oThTitled

\begin{proof}

We show SN of $\RR = \Ax\uplus\Bx$ by using Prop. \ref{th:pred-lab}.
Suppose that \QQ is SN by the General Schema
with a well-founded order $\gt_\SigB$.
We define an order on \Sig by
\begin{equation}
  \begin{array}[h]{llll}
h \lgt k \lgt &f_v \lgt f_w,\; g_w
                  \lgt c,\; <-,->_b,\; \bot_b
  \end{array}
\label{eq:prec}
\end{equation}
\begin{itemize} \item for all $h,k \in \QQsig$ with $ h \gt_\SigB k$,
\item for all $f,g\in\RRRsig,\, v,w\in\TsigV$ 
with $v \;\AprojSUB\; w$ and
$f$ is the head of the lhs of a $\Ax$-rule and $g$ appears in the
corresponding rhs (possibly $g=f$),
\item for all $c\in \OH,\; b \in \BB$.
\end{itemize}
The order $\gt_\Sig$ is well-founded
because $\gt_\SigB$ is well-founded.
We show $\labed\RR \uplus \Decl 
=\labed\RRR \uplus \labed\QQ \uplus \Decl$
satisfies \GS with $\gt_\Sig$.
\begin{enumerate}[label={{\arabic*.}},leftmargin=*]

\item $\labed\RRR$ satisfies \GS.
Take a labelled rule $f_v(\vec{\vec x.t})\TO r \in \labed\Ax$. 

\noindent
We show that for all $r' \tle r$,
$\CCl \ni r'$ by induction on the structure of $r'$. \begin{itemize}[leftmargin=1.2em,noitemsep]\label{itm:Acase}
\item Case $r'=M[\vec x]$. By assumption, $M$ is accessible in some of $\vec t$.
Variables $\vec x \in \CCl$. Therefore, \textbf{(meta $\va m$)} is applied,
we have $M[\vec x] \in \CCl$.
\item Case $r'$ is a variable $x$. Then $\CCl \ni x$.

\item Case $r' = g_w({\vec{\vec y.s}})$ with $g\in\SigA$ (possibly $g=f$). 
Since $r'$ is a sub-meta-term of the rhs $r$ of  a $\labed\Ax$-rule,
\begin{meqa}
v = \ext\phi(f(\vec{\vec x.\forget t}))
\;\;(\toAproj\!\!\union\toAproj\!\!\!\!\!\!\!\circ\,\tgtS)\;\;
\ext\phi(g(\vec{\vec y.\forget s})) = w\end{meqa}
where the term $\forget t$ is obtained by
deleting all labels in a labelled term $t$.
Then we have $f_v  \gt_\Sig g_w$.
By I.H., $\CCl \ni \vec s$. 
Applying \textbf{(fun $f_v \gt_\Sig g_w$)},
$\CCl \ni r'$. \label{itm:gw-case}

\item Case $r' = c({\vec{\vec y.s}})$ with $c\in\OH$.
Then $f_v  \gt_\Sig c$. 
By I.H., $\CCl \ni \vec{s}$, hence 
\textbf{(fun $f_v \gt_\Sig c$)}
is applied, $\CCl \ni r'$.

\end{itemize}
Therefore, $\CCl \ni r$.

\item ${\labed\QQ}$ satisfies \GS with $\gt_\Sig$. In $\labed\Bx$, 
any \SigB-symbol is not labelled
and any labelled \SigA-symbol is smaller 
(w.r.t. $\lt_\Sig$) than a \SigB-symbol.
These facts and the assumptions \ref{itm:I-A-SN} and \ref{itm:I-B-SN}
imply the desired result.

\item $\Decl$ satisfies \GS:  We need to show that each rule
$$
f_v(\vec{x_1}.M_1[\vec{x_1}],\ooo,\vec{x_n}.M_n[\vec{x_n}]) \;\TO\;
f_w(\vec{x_1}.M_1[\vec{x_1}],\ooo,\vec{x_n}.M_n[\vec{x_n}])
$$
with $v \;\AprojSUB\; w$
satisfies \GS. This holds because \GS checks the root symbols 
by \textbf{(fun $f_v \gt_\Sig g_w$)} 
for $f,g\in\SigA$, and
each metavariable is accessible.

\end{enumerate}
Therefore, $\labed{\RR}\uplus \Decl$ satisfies \GS, hence SN 
by Proposition \ref{th:lGS}.
Finally, applying Prop. \ref{th:pred-lab}, we conclude that \RR is SN.
\QED

\subsection{Variations}

\subsubsection{\GS with other term orders}
In~\cite{Blanqui-TCS}, a general and comprehensive account of \GS is presented, where 
a preorder on function symbols and a preorder to compare the arguments of 
equivalent function symbols are abstracted to the notion of
\W{valid (status) \FF-quasi-ordering}~\cite[Def.7,9]{Blanqui-TCS}.
\GS reviewed in \Sec \ref{sec:GS} is
merely one instance of it.
Importantly, Thm. \ref{th:modI} holds for any variation of \GS
because the proof does not depended on any particular valid \FF-quasi-ordering.
These changes can enhance applicability of the theorem.

\Cor
Thm. \ref{th:modI} holds by changing \PGS to 
another version of the General Schema using a different valid \FF-quasi-ordering.
For instance, the use of
\begin{itemize*}[label={}]
\item covered-subterm ordering~\cite[Def.5]{IDTS}, or
\item structural subterm ordering~\cite[Def.13]{Blanqui-TCS}
\end{itemize*}
is possible.
\oCor

For instance, \GS with the structural subterm ordering can check SN of a rule
\begin{solLmath}
\s{lim(x.M[x])+Y $\TO$ lim(x.M[x]+Y)}   
\end{solLmath}
\cite[Sec. 4.6]{Blanqui-MSCS}
for addition of ordinals, while the stable subterm ordering cannot.
Changing \PGS, \RR can involve this kind of rules.

Similarly, \PGS might be changed to the
computability path ordering (CPO) with accessible subterms 
\cite[Sec.7.2]{Blanqui-comp} because the proof of Thm. \ref{th:modI}
only uses the features of \GS on comparison of function symbols and accessible variables,
and CPO with accessible subterms also has them.
We will pursue these directions elsewhere
and will clarify detailed conditions on these variations.

\subsubsection{Polymorphic and dependently typed cases}
Furthermore, changing the General Schema to GS defined in~\cite{Blanqui-MSCS},
and generalising the \resyss to the polymorphic \resyss developed 
in~\cite{pocr},
we can obtain a polymorphic version of Thm. \ref{th:modI}.
This setting is more natural than the present molecular typed rules
(see the discussion on difference between molecular types and polymorphic
types in~\cite{pocr}) in practical examples
because we do not need to check parameterised instances 
of function symbols and rules as in Example \ref{ex:SigStl}.
But since in the framework of~\cite{Blanqui-MSCS}, 
the lhss of rules cannot involve binders or HO patterns,
we need to restrict the form of lhss to apply
this version of \GS.
For the dependently typed case, 
the recent development~\cite{lmdPi-dp} will also be useful.

 \section{Application 1: Algebraic Effect Handlers and Effect Theory}
\label{sec:eff}

As an application, we demonstrate that our theorem
is useful to prove the termination of a calculus with algebraic effects.
The background of this section is as follows.
Plotkin and Power introduced
the \W{algebraic theory of effects} 
to axiomatise various computational effects~\cite{PlotPowerFos}
and they correspond to computational monads
of Moggi~\cite{comp-lambda-monad}.
Plotkin and Pretnar formulated
\W{algebraic effects and handlers}~\cite{handlingEff}, which provided an
alternative to monads as a basis for effectful programming
across a variety of functional programming languages.

We prove termination of 
an effectful \lmd-calculus with {effect handlers}
that respects an {effect algebraic theory}.  
First, we formulate the \W{multi-adjunctive metalanguage} (\MAM),
for 
an effectful \lmd-calculus (\Sec \ref{sec:mam}).
Secondly, we provide an effect theory (\Sec \ref{sec:eff-terms}).
Thirdly, we give an effect handler (\Sec \ref{sec:handler}).
Two novelties arise in this section.
\begin{enumerate}
\item We use a single framework to formalise \MAM, 
an effectful handler, and an effect theory:
they form second-order \resyss.
\item We prove the termination of the combination of
these by the modular termination theorem: Thm. \ref{th:modI}.
\end{enumerate}

\subsection{Type constructors}\label{sec:typeConstr}
We first define type constructors for an effectful calculus.
\begin{itemize}[noitemsep]
\item \tyN{atomic type}{Unit}{unit type}{()}
\item \tyN{binary}{Pair}{product type}{a_1\X a_2}
\item \tyN{binary}{Sum}{sum type}{a_1+ a_2}
\item \tyN{binary}{CPair}{product of computation types}{a_1\& a_2}
\item \tyN{unary}{U}{thunked type}{U_E\; a}
\item \tyN{unary}{F}{computation type}{F\; a}
\item \tyN{binary}{Arr}{arrow type}{a_1\to a_2}
\end{itemize}

\subsection{A calculus for algebraic effects}\label{sec:mam}
We use the core calculus \MAM for effectful computation given in~\cite{FosterKammerEffJ},
which is an extension of 
Levy's call-by-push-value (CBPV) calculus~\cite{CBPV}.
We formulate \MAM as a second-order \resys $(\SigMAM\uni\OH,\s{MAM})$.
The signature \SigMAM consists of the following defined function symbols
\begin{solM}
  bang  : U(c) $\to$ c
  caseP : Pair(a1,a2),(a1,a2 $\to$ c) $\to$ c
  case  : Sum(a1,a2),(a1 $\to$ c),(a2 $\to$ c) $\to$ c
  let : F(a),(a $\to$ c) $\to$ c
  app : Arr(a,c),a $\to$ c
  prj1 : CPair(c1,c2) $\to$ c1     ; prj2 : CPair(c1,c2) $\to$ c2
\end{solM}
and the set of constructors \OH consists of
\begin{solM}
  unit  : Unit             ; pair : a1,a2 $\to$ Pair(a1,a2)
  inj1  : a1 $\to$ Sum(a1,a2) ; inj2 : a2 $\to$ Sum(a1,a2)    
  cpair : c1,c2 $\to$ CPair(c1,c2)
  thunk : c $\to$ U(c)
  return : a $\to$ F(a)
  lam : (a $\to$ b) $\to$ Arr(a,b)
\end{solM}

The set \s{MAM} of \MAM's computation rules is given by\footnote{\code{t@s} is the abbreviation of 
\code{app(t,s)}.}

\begin{solM}
  (beta)  lam(x.M[x])@V                     $\TO$ M[V]
  (u)     bang(thunk(M))                    $\TO$ M
  (prod1) prj1(cpair(M1,M2))                $\TO$ M1
  (prod2) prj2(cpair(M1,M2))                $\TO$ M2
  (caseP) caseP(pair(V1,V2),x1.x2.M[x1,x2]) $\TO$ M[V1,V2]
  (case1) case(inj1(V),x.M1[x],y.M2[y])     $\TO$ M1[V]
  (case2) case(inj2(V),x.M1[x],y.M2[y])     $\TO$ M2[V] 
  (f)     let(return(V),x.M[x])             $\TO$ M[V]
\end{solM}
Using this formulation, SN of \s{MAM} is immediate
because it satisfies \GS. More precisely,
every rhs of \s{MAM} involves no function symbol and
every metavariable in \s{MAM} is accessible.
The original paper proposing \MAM~\cite[Thm. 2]{FosterKammerEffJ} 
described a sketch of proof of SN of \MAM, but 
the details were omitted.
Our formulation by a \resys is generic and 
the system \SOL automatically proves SN of \verb!0Ex52_MAM.hs! in \fn{arXiv'20}
of the \SOL web interface~\cite{SOLweb}.

Note that \SOL merely checks SN of the finite number of function symbols
and rules defined in the file \verb!0Ex52_MAM.hs!.
To conclude actual SN of the \resys~\MAM,
we use meta-theoretic reasoning. 
Looking at the output of checking process of \GS,
we see that all the accessibility conditions are satisfied even if we
replace the type letters \s{a,a1,a2,c,c1,c2} used in the signature with 
concrete mol types
generated by the type constructors defined in \Sec \ref{sec:typeConstr},
because there are no constructors violating the positivity condition 
(written as ``\s{is positive}'' in \SOL's output)
\ref{itm:acc3} of the accessibility predicate in Def. \ref{def:acc}.
Therefore, the infinite number of \MAM' computation rules 
obtained by instantiating
the type letters \s{a,a1,a2,c,c1,c2} with concrete types satisfy \GS.

\subsection{Effect theory}\label{sec:eff-terms}

Next we extend \s{MAM} to have
an effect handler with effect theory. 
To keep the discussion simple, we consider a particular theory, i.e.
the theory of global state~\cite{PlotPowerFos,SamTwoCotensor}
for a single location. We take the type \s{N} of natural numbers for the state.
We define the signature $\SigGl$ by
\begin{solM}
  get : (N $\to$ F(N)) $\to$ F(N)
  put : N,F(N) $\to$ F(N)
  sub : (N $\to$ F(N)), N $\to$ F(N)
\end{solM}
It consists of the operations \code{get(v.t)} (looking-up the state, binding the 
value to \code{v}, and continuing \code{t}) and
\code{put(v,t)} (updating the state to \code{v} and continuing \code{t}), 
and the substitution operation \code{sub(x.t,s)} that replaces \s{x} in \s{t}
with \s{s}.
The theory of global state~\cite{PlotPowerFos,Substituion-Jumps} 
can be stated as a \resys~$(\SigGl\uni\set{\s{return}},\s{gstate})$ defined by

\begin{solM}
 (lu)   get(v.put(v,X))        $\TO$ X
 (ll)   get(w.get(v.X[v,w]))   $\TO$ get(v.X[v,v])
 (uu)   put(V,put(W,X))        $\TO$ put(W,X)
 (ul)   put(V,get(w.X[w]))     $\TO$ put(V,sub(w.X[w],V))
 (sub1) sub(x.return(x),K)     $\TO$ return(K)
 (sub2) sub(x.M,K)             $\TO$ M  
 (sub3) sub(x.get(v.M[v,x]),K) $\TO$ get(v.sub(x.M[v,x],K))
 (sub4) sub(x.put(V,M[x]),K)   $\TO$ put(V,sub(x.M[x],K))
\end{solM}
These axioms have intuitive reading. For example,
the axiom \code{(lu)} says that looking-up the state, binding the value to \code{v}, then
updating the state to \code{v}, is equivalent to doing nothing.
The axiom \code{(ul)} says that updating the state to \code V, then
looking-up and continuing \code X with the looked-up value,
is equivalent to updating the state to \code V and continuing \code X 
with \code V.

Plotkin and Power showed that the monad corresponding to 
the theory of global state (of finitely many locations) is the state monad
~\cite{PlotPowerFos}.

Crucially, $(\SigGl\uni\set{\s{return}},\s{gstate})$ 
does not satisfy \GS.
\TGS checks that a recursive call at the rhs must be 
with a strict sub-meta-term of an argument at the lhs. 
In case of \code{(ll)}, the recursive call of \code{get} happens with 
\code{v.X[v,v]}, which is not a sub-meta-term of 
\code{w.get(v.X[v,w])} at the lhs.
Moreover, \code{(ul)} requires a precedence $\s{put}\gt\s{sub}$ while
\code{(sub4)} requires $\s{sub}\gt\s{put}$, which violates
well-foundedness.

Using a different method, the \resys~$(\SigGl\uni\set{\s{return}},\s{gstate})$ is shown to be SN.
We count the number of symbols \s{get,put,sub} using
the weights defined by
\[
  \den{\s{put}}(v,x) = x+1\quad
  \den{\s{get}}(x) = 2x + 2\quad
  \den{\s{sub}}(x,v) = 2x +1
\]
In each rule, the weights are decreasing such as
\begin{meqa}
  \s{(ul)}\qquad (2x+2)+1 &\gt (2x+1) +1\\
  \s{(sub4)}\quad 2(x+1)+1 &\gt (2x+1)+1
\end{meqa}
therefore it is SN.
Note that since $\s{N}\not=\s{F(N)}$ and there is no function symbol of
type $\s{F(N)}\to\s{N}$ in $\SigGl$, 
the function symbols \s{get,put,sub} cannot occur in the first argument of \s{put} and
the second argument of \s{sub} as an instance of the argument \s{V}.
Therefore, the parameter $v$ is not used in the weights.

\subsection{Effect handler}\label{sec:handler}

\W{An effect handler} 
provides an implementation of effects by
interpreting algebraic effects as actual effects.
The handler~\cite{HandlerAction,FosterKammerEffJ}
for effect terms for global states 
can be formulated as a \resys $(\SigGl\uni\SigMAM\uni\SigHandle,\s{Handle})$ 
as follows.
\begin{solM}
  handler : (N $\to$ F(N)), (Arr(N,F(N)) $\to$ F(N)), (N,F(N) $\to$ F(N)), F(N) $\to$ F(N)
\end{solM}

\begin{solM}
 (h\_r) handler(RET,GET,PUT,return(X))  $\TO$ RET[X]
 (h\_g) handler(RET,GET,PUT,get(x.M[x]))$\TO$ GET[  lam(x.handler(RET,GET,PUT,M[x]))]
 (h\_p) handler(RET,GET,PUT,put(P,M))   $\TO$ PUT[P,lam(x.handler(RET,GET,PUT,M))]
\end{solM}
Note that \s{RET},\s{GET},\s{PUT} are metavariables.
For brevity, the first three arguments of \s{handler} are abbreviated,
and they are formally \eta-expanded forms:
\begin{center}
\s{handler(y.RET[y],k.GET[k],p.k.PUT[p,k],$\ccc$)}.  
\end{center}
We consider a standard interpretation of global states 
using parameter-passing~\cite{handlingEff,HandlerAction,FosterKammerEffJ}.
Taking states and values to be \s{N}, we have the computation type 
$\s{F(N)} = \s{Arr(N,N)}$.
Then running the handler can be given by

\begin{solM}
 runState : Arr(N,N) $\to$ Arr(N,N)
 (run) runState(t) $\TO$ handler(y.lam(z.y), k.lam(n.(k@n)@n), p.k.lam(n.k@p), t)
\end{solM}
This interprets \s{return,get,put} in
an effect term \s{t} as the corresponding arguments of \s{handler},
which are interpretation of global state in the state-passing style.

The \resys $(\SigGl\uni\SigMAM\uni\SigHandle,\s{Handle})$ is 
immediately shown to be SN by \GS. \SOL automatically proves it
(try \verb!0Ex54_Handle.hs! in \fn{arXiv'20} of the SOL web interface~\cite{SOLweb}).
From it, we can conclude actual SN of \s{Handle} by using meta-theoretic reasoning
as mentioned in \Sec \ref{sec:mam}.

\subsec{Proof of SN}
An effect term expresses an effectful program. For example,
\[
tm \;\deq\;\, \s{get(x.put(inc(x),get(y.put(y,get(z.return(z))))))}
\]
expresses the imperative program~\cite{HandlerAction}
\begin{solM}
  x := get; put inc(x); y:= get; put y; z := get; return z
\end{solM}
where $\s{inc} : \s{N} \to \s{N} \in \SigHandle$ is intended to be the increment operation.
If the initial store is set to \s{0}, then clearly this program returns
\s{inc(0)}.
Formally, it is computed using the term $tm$ with the handler as
\begin{solM}
 runState($tm$)@0
 $\TO$ handler(y.lam(z.y), k.lam(n.(k@n)@n), p.k.lam(n.k@p),$tm$)@0
 $\TOO$ lam(z.inc(z))@0 $\TO$ inc(0)
\end{solM}
This can be computed using $\s{MAM}\uni\s{Handle}$, or
$\s{gstate}\uni\s{MAM}\uni\s{Handle}$. The latter is more efficient than the former,
because \s{gstate} expresses program equivalences and the application
of them to an effect term optimizes the program.

Now, we consider the main problem of this section: SN of the whole \resys 
\begin{equation}
(\SigGl\uni\SigMAM\uni\SigHandle\uni\OH,\;\s{gstate}\uni\s{MAM}\uni\s{Handle})  
\label{eq:eff}
\end{equation}
\TGS does not work to show SN of it because 
\s{gstate} does not satisfy \GS. Therefore, we divide it into
\begin{meqa}
(\SigA\uni\OH,\Ax) &= (\SigGl\uni\OH,\;\s{gstate}),
\\
(\SigA\uni\SigB\uni\OH,\Bx) &= (\SigGl\uni(\SigMAM\uni\SigHandle)\uni\OH,\;\s{MAM}\uni\s{Handle}).
\end{meqa}
The \resys (\ref{eq:eff}) is not a disjoint union of
\Ax and \Bx,
and is actually a \W{hierarchical combination} (cf. Remark \ref{rem:ass})
that \W{shares constructors}
\OH. 
The lhss of \s{Handle} ($\subseteq\Bx$) involve
defined function symbols \s{get},\s{put} in \SigA.

We apply the modularity Thm. \ref{th:modI}.
Assumption \ref{def:setting} is satisfied
because the \resys (\ref{eq:eff}) is finitely branching and satisfies the 
the \SigA-layer
condition. We check the assumptions. We define the well-founded order 
on types by
\[
T(a_1,\ooo,a,\ooo,a_n) \gt_\TT a 
\]
for every $n$-ary type constructor $T$, and every type $a$,
where the lhs's $a$ is placed at the $i$-th argument
of $T$ for every $i=1,\ooo,n$.

\begin{enumerate}
\item $\Ax = \s{gstate}$ is accessible. This is immediate because
 the crucial case \ref{itm:acc3} in Def. \ref{def:acc} checks
the type comparison for the arguments of 
\s{get},\s{put},\s{return}, which holds by $\s{N}\lt_\TT \s{F(N)}$.

\item $(\SigA\uni\OH\uni\Sig_\Proj,\s{gstate}\uni\Proj)$  is SN. This is again
established by the weights given in \Sec \ref{sec:eff-terms}.

\item $(\SigA\uni\SigB\uni\OH,\;\s{MAM}\uni\s{Handle})$ is SN by \GS. 
This is immediate by 
applying \GS with the precedence
\[
\s{handler} \gt_\Sigma \s{lam}
\]
to the \resys.
The rhss of \s{MAM} involve no function symbols and 
the rhss of \s{Handle} involve \s{handler},\s{lam}. 
To check that each recursive call of
\s{handler} happens with a smaller argument, 
here we use the structural subterm ordering~\cite[Def.13]{Blanqui-TCS}
to establish that \s{M[x]} is smaller than \s{get(x.M[x])}.
Every metavariable
is accessible.
\end{enumerate}
Hence we conclude that the \resys (\ref{eq:eff}) is SN.

This termination result is general. 
Although in this section, we consider a
particular handler for \s{get} and \s{put},
any effect handler is shown to be SN by \GS along this way.
\MAM is SN regardless of used effects.
To the best of our knowledge, this is the first report on 
how to prove the termination of the combination of 
an effectful \lmd-calculus, an effect handler
and effect theory.

 \section{Application 2:  Splitting a System into FO and HO parts}\label{ex:fo-split}
If a \resys contains first-order computation rules,
it is often useful to split the system
into the higher-order part and the first-order part
to which we can apply a powerful termination checker for
first-order term rewriting systems, such as 
\fn{AProVE}~\cite{AProVE}. 
The Modular Termination Theorem (Thm. \ref{th:modI}) can be used for this purpose. 

A FO \resys $(\Sig,\RR)$ is a \resys 
if the type of every function symbol $f\in\RR$ is of the form
$f:a_1,\ooo,a_n \to b$, where $a_i$'s and $b$ are mol types. 
Suppose that a \resys \RR can be split as $\RR=\Ax\uni\Bx$,
where \Ax is a set of FO rules without \SigB-symbols, \Bx is a set of second-order rules
and Assumption \ref{def:setting} is satisfied.
For every FO \resys, 
the rhss of rules are second-order patterns because no meta-application
exists in the arguments of a function symbol.
Hence if the FO \resys \Ax is accessible,
\Ax with \Proj is SN (using any method), and \Bx is SN by GS,
then we can conclude that $\Ax\uni\Bx$ is SN by Thm. \ref{th:modI}.

\subsec{Implementation}\label{sec:web}
We have implemented this FO splitting method
by extending the tool \textsf{SOL}~\cite{SOL}.
We will also refer to this extended version as \SOLall.
The system \SOLall consists of about 8000 line Haskell codes.
The web interface of \SOLall is available at the author's homepage.

\subsec{Benchmark}\label{sec:bench}
The Termination Problem Database (TPDB)\footnote{\url{http://termination-portal.org/wiki/TPDB}}
stores a collection of various rewrite systems for termination.
In the TPDB,
``\verb!Applicative_first_order!'' problems 
and some other problems in the higher-order category
contain
such examples, which are a mixture of difficult FO systems and 
HO systems.
To show effectiveness of this method, 
we selected 62 problems from TPDB
and did a benchmark to solve these problems
by the previously proposed system \SOL (2017)~\cite{SOL} and the extended
\SOLall (current version) for comparison.
We conducted the benchmark on a machine with
Intel(R) Xeon E7-4809, 2.00GHz 4CPU (8core each), 
256GB memory, Red Hat Enterprise Linux 7.3, and set timeout 400 seconds.

\SOL (2017) solved 31 problems, and
\SOLall (current version) solved 57 problems out of 62,
which clearly improves \SOL (2017). \SOLall (current version) 
could solve 26 more problems
than \SOL (2017).
The output details are available\footnote{\url{http://solweb.mydns.jp/bench/}}
and shown in Table \ref{fig:table} in the Appendix,
where the use of a modularity method is indicated at the final column.

\begin{figure}\small

\begin{minipage}[t]{\textwidth}
\begin{wrapfigure}[8]{r}{20em}\y{-2em}\scriptsize
\begin{tabular}{|c|>{\columncolor[rgb]{0.8, 0.8, 0.8}[\tabcolsep]}c|c|c|c|} \hline
Result & 
{\SOLall} &
{\;\;\WANDA\;\;} &
{\!\scriptsize \SizeChange\!}
\\
\hline
YES &   199&     151&     93     \\
NO  &20&          14&    0      \\
MAYBE &  37&      28&     163      \\
error & 5&      66&      3      \\
timeout & 0 & 2 & 2\\
\hline
\hline
Score & 219 & 165 & 93 \\
\hline
\end{tabular}
\end{wrapfigure}
The table shows the number of judged results by each tool.
The result YES/NO is a judged result of SN.
MAYBE is a result when the tool cannot judge.
The score of a tool is the number of judged answers.
(\SOLall is named as ``\textsf{sol 37957}'')

\bigskip\bigskip

{
{ 
Score:\;\;\; \url{http://group-mmm.org/termination/competitions/Y2018/}}
\\Details:\; {\scriptsize \url{http://group-mmm.org/termination/competitions/Y2018/caches/termination_30047.html} }
\\Termination Competition Official:\\ { \url{http://termination-portal.org/wiki/Termination_Competition_2018}}

and also proceed [Status]-[Results]
}

\medskip
{\small
Note that there is one conflict of judgement
in the problem {\scriptsize \s{Hamana\_17/restriction.xml}} 
(which swaps two \s{new}-binders and is a problem the present author submitted)
due to different interpretations of the higher-order rule format among
the tools.
}
\end{minipage}

\caption{Results and scores in Termination Competition 2018}\label{fig:score-table}
\end{figure}

 \subsec{Example}\label{sec:termcomp}
We show one of these problems:
\verb!Applicative_05__mapDivMinusHard!.

\begin{solT}
 (1)  map(x.F[x],nil)      $\TO$ nil
 (2)  map(x.Z[x],cons(U,V))$\TO$ cons(Z[U],map(x.Z[x],V))
 (3)  minus(W,0)           $\TO$ W
 (4)  minus(s(P),s(X1))    $\TO$ minus(p(s(P)),p(s(X1)))
 (5)  p(s(Y1))             $\TO$ Y1
 (6)  div(0,s(U1))         $\TO$ 0
 (7)  div(s(V1),s(W1))     $\TO$ s(div(minus(V1,W1),s(W1)))
\end{solT}

This \s{mapDivMinusHard} does not satisfy \GS because of \s{(7)}, which
is not structural recursive.
\SOLall splits it into the FO part
$\Ax=\set{\s{(3)}\RM{-}\s{(7)}}$
and HO part $\Bx=\set{\s{(1)}\RM{-}\s{(2)}}$.
Then \Ax with \Proj can be proved to be SN by an external FO termination checker and
\Bx satisfies \GS. 
Then \SOLall concludes that \s{mapDivMinusHard} is SN by Thm. \ref{th:modI}.

As a more comprehensive evaluation,
\SOLall
participated to the higher-order union beta category of
the International Termination Competition 2018
held at the Federated Logic Conference (FLoC 2018) in Oxford,
U.K.
This event was a competition for the number of checked results
of termination problems by automatic tools.
In the higher-order category, 263 problems of higher-order rewrite systems
taken from TPDB
were given. 
Three tools participated in the higher-order category:
WANDA~\cite{WANDA}, SizeChangeTool~\cite{lmdPi-dp}, and \SOLall.
\SOLall judged the greatest number of problems
among three tools as shown in Fig. \ref{fig:score-table}.
A main reason derives from the fact that \SOLall has 
a modular SN checking method based on Theorem \ref{th:modI}.

 \section{Related Work and Summary}
\label{sec:related}
\subsection{Related work}
A rich body of work describes termination and modularity of
first-order and higher-order rewrite systems including 
\cite{Zantema,Ohlebusch,dep,moduldar-dp,IDTS00,IDTS,HORPO-JACM,
Kusakari18,WANDA,size-change,Blanqui-TCS,Toyama-NTT,Ohlebusch,moduldar-dp,Gramlich,cube}.
The modular termination theorem of the form of Thm. \ref{th:modI},
i.e., 
the combination of (restricted classes of) 
two second-order computation systems with shared constructors,
has not been reported in the literature to date.
\cite{harness} provides a dependency pair method to split a HO termination problem
into FO and HO parts, which might be regarded as a modularity of SN for
the combination of  FO and HO rewrite systems.
An important difference is that 
our result covers strictly more than the combination
of FO and HO systems as described in \Sec\ref{ex:fo-split}.

Originally,
our proof of modularity was inspired 
by a method of proving modularity of SN
for a rewrite system \calR and a recursive 
program scheme \PP by semantic labelling
(\cite{self} for the FO case, and~\cite{HSL} for the second-order case).
In this method, the normal forms computed by \PP
are taken as the ``semantics'', and are used
as labels to show termination of $\calR\uni\PP$, where SN of \PP is crucially
important to determine the semantics.

For the present work, 
we chose \SigA-terms
computed by a \SigA-substitution $\aext\phi$ as the ``semantics''.
Also, they are used as labels to show termination of $\Ax\uni\Bx$, where SN of \Ax 
is crucially important to determine the semantics\footnote{It forms a
  quasi-model of \Ax consisting of 
\Sig-terms with a well-founded preorder $\preAprojSUB$.}.

\smallskip

The dependency pair (DP) method is a successful method of proving
termination of various first-order~\cite{Arts-dep,dep} and
higher-order rewrite systems 
\cite{Kusakari18,WANDA,lmdPi-dp}.
One may find that several similarities exist between the DP method and the
present proof strategy.
A reason might be that the DP method is a general method to prove
termination by conducting modular analysis of the corresponding dependency pairs 
\cite{moduldar-dp}.

More fundamentally, a similarity between DP and the present work 
can be found at the foundational level.
At the early stage~\cite{Arts-dep,dep-cs},
the DP technique was \W{based on semantic labelling},
where the normal forms by a ground-convergent rewrite system \EE
\footnote{A target rewrite system \calR to prove SN is 
assumed to be contained in \EE.}
are taken as the ``semantics'',
and are used as labels of rules to establish the DP method.
The ground convergence of \EE is crucially important to determine the semantics.
In the later refined DP method,
the use of that semantic labelling was dropped~\cite{dep}.
But this fact illustrates that semantic labelling is a natural starting point 
to tackle the modularity problem of termination.

Our modularity result and its proof might be reformulated 
using the DP method, where
the \Aproj-part might be regarded as usable rules.
However, the static DP method 
\cite{Kusakari18} 
might be insufficient to simulate our theorem completely.
A mismatch seems to exist between SN of higher-order rules
and non-loopingness of the corresponding dependency pairs. 
For example, while the second-order computation rule \s{f(0)$\TO$g(x.f(x))} is terminating 
(and accessible, the lhs is a pattern),
the corresponding dependency pair \s{f(0)$\to$f(x)} is looping.
Therefore, one cannot replace
our assumption of SN of \Aproj completely with
non-loopingness of the corresponding DPs.
But employing some ideas of the DP method
might increase the power of Thm. \ref{th:modI},
which is left as a future work.

\subsection{Summary}
We have presented a new modular proof method of termination for second-order
computation. The proof method is useful
for proving termination of higher-order foundational calculi. To establish
the method, we have used a version of the semantic labelling translation and
Blanqui's General Schema: a syntactic criterion of strong normalisation. As
an application, we have applied this method to show termination of a variant of
call-by-push-value calculus with algebraic effects, an effect handler and
effect theory.  We
have also shown that our tool \SOLall is effective to solve higher-order
termination problems.

 \bibliographystyle{alphaurl}
\bibliography{bib}
\newpage
\begin{figure}[H]
{\scriptsize
 \tt \begin{center}
\begin{tabular}{|l|c|c||c|c|c|c|c|c|} \hline
  \renewcommand{\arraystretch}{1.1}
{Problem} & 
\multicolumn{2}{l}{\SOL{\scriptsize (2017)} (sec.)}|& 
\multicolumn{2}{l}{\SOLall \;(sec.)}|&\; Used method
 \\
\hline
Applicative\_05\_\_Ex3Lists & YES & 0.021 & YES & 0.022 &  \\
Applicative\_05\_\_Ex4MapList & YES & 0.019 & YES & 0.019 &  \\
Applicative\_05\_\_Ex5Folding & YES & 0.025 & YES & 0.022 &  \\
Applicative\_05\_\_Ex5Sorting & YES & 0.025 & YES & 0.025 &  \\
Applicative\_05\_\_Ex6Folding & YES & 0.024 & YES & 0.018 &  \\
Applicative\_05\_\_Ex6Recursor & YES & 0.019 & YES & 0.018 &  \\
Applicative\_05\_\_Ex7Sorting & YES & 0.024 & YES & 0.025 &  \\
Applicative\_05\_\_Ex7\_9 & \;MAYBE\; & 0.025 & MAYBE & 0.025 &  \\
Applicative\_05\_\_Ex9Maps & YES & 0.022 & YES & 0.019 &  \\
Applicative\_05\_\_Hamming & MAYBE & 0.035 & NO & 0.044 &  \\
Applicative\_05\_\_mapDivMinus & MAYBE & 0.027 & YES & 0.101 & modularity \\
Applicative\_05\_\_mapDivMinusHard & MAYBE & 0.022 & YES & 0.12 & modularity \\
Applicative\_05\_\_termMonTypes & MAYBE & 0.021 & NO & 0.016 &  \\
Applicative\_first\_order\_05\_\_\#3.16 & YES & 0.023 & YES & 0.025 &  \\
Applicative\_first\_order\_05\_\_\#3.18 & MAYBE & 0.03 & YES & 0.142 & modularity \\
Applicative\_first\_order\_05\_\_\#3.2 & MAYBE & 0.023 & YES & 0.102 & modularity \\
Applicative\_first\_order\_05\_\_\#3.22 & MAYBE & 0.022 & YES & 0.107 & modularity \\
Applicative\_first\_order\_05\_\_\#3.25 & MAYBE & 0.02 & YES & 0.114 & modularity \\
Applicative\_first\_order\_05\_\_\#3.27 & YES & 0.022 & YES & 0.021 &  \\
Applicative\_first\_order\_05\_\_\#3.32 & MAYBE & 0.021 & YES & 0.095 & modularity \\
Applicative\_first\_order\_05\_\_\#3.36 & YES & 0.025 & YES & 0.026 &  \\
Applicative\_first\_order\_05\_\_\#3.38 & YES & 0.028 & YES & 0.035 &  \\
Applicative\_first\_order\_05\_\_\#3.40 & MAYBE & 0.025 & YES & 5.061 &  \\
Applicative\_first\_order\_05\_\_\#3.45 & MAYBE & 0.022 & YES & 0.099 & modularity \\
Applicative\_first\_order\_05\_\_\#3.48 & MAYBE & 0.025 & YES & 0.119 & modularity \\
Applicative\_first\_order\_05\_\_\#3.52 & MAYBE & 0.021 & YES & 0.093 & modularity \\
Applicative\_first\_order\_05\_\_\#3.55 & MAYBE & 0.028 & YES & 0.221 & modularity \\
Applicative\_first\_order\_05\_\_\#3.57 & MAYBE & 0.026 & YES & 0.17 & modularity \\
Applicative\_first\_order\_05\_\_\#3.6 & YES & 0.027 & YES & 0.032 &  \\
Applicative\_first\_order\_05\_\_\#3.8 & MAYBE & 0.023 & YES & 0.149 & modularity \\
Applicative\_first\_order\_05\_\_01 & YES & 0.025 & YES & 0.026 &  \\
Applicative\_first\_order\_05\_\_02 & YES & 0.024 & YES & 0.025 &  \\
Applicative\_first\_order\_05\_\_06 & YES & 0.021 & YES & 0.021 &  \\
Applicative\_first\_order\_05\_\_08 & YES & 0.022 & YES & 0.028 &  \\
Applicative\_first\_order\_05\_\_11 & YES & 0.028 & YES & 0.047 &  \\
Applicative\_first\_order\_05\_\_12 & YES & 0.026 & YES & 0.029 &  \\
Applicative\_first\_order\_05\_\_13 & MAYBE & 0.023 & YES & 17.03 &  \\
Applicative\_first\_order\_05\_\_17 & YES & 0.024 & YES & 0.027 &  \\
Applicative\_first\_order\_05\_\_18 & YES & 0.022 & YES & 0.022 &  \\
Applicative\_first\_order\_05\_\_21 & MAYBE & 0.022 & YES & 0.128 & modularity \\
Applicative\_first\_order\_05\_\_29 & YES & 0.025 & YES & 0.025 &  \\
Applicative\_first\_order\_05\_\_30 & MAYBE & 0.022 & YES & 0.313 & modularity \\
Applicative\_first\_order\_05\_\_31 & MAYBE & 0.024 & MAYBE & 0.024 &  \\
Applicative\_first\_order\_05\_\_33 & MAYBE & 0.023 & TIMEOUT & 400.003 &  \\
Applicative\_first\_order\_05\_\_hydra & YES & 0.024 & YES & 0.025 &  \\
Applicative\_first\_order\_05\_\_minsort & MAYBE & 0.028 & TIMEOUT & 400.005 &  \\
Applicative\_first\_order\_05\_\_motivation & MAYBE & 0.022 & YES & 0.084 & modularity \\
Applicative\_first\_order\_05\_\_perfect & MAYBE & 0.024 & YES & 0.122 & modularity \\
Applicative\_first\_order\_05\_\_perfect2 & MAYBE & 0.03 & YES & 0.152 & modularity \\
h21 & MAYBE & 0.023 & YES & 0.249 &  \\
h22 & MAYBE & 0.026 & YES & 0.337 & modularity \\
h23 & MAYBE & 0.027 & YES & 0.376 & modularity \\

\hline
\end{tabular}
\end{center}
\y{-1em}
\rm\normalsize
\caption{Appendix: Comparison of \SOL (2017)\cite{SOL} and \SOLall (this paper)
}
\label{fig:table}
}
\end{figure}

 \section*{Appendix}
\appendix
 
\section{The General Schema for Labelled Computation Systems}\label{sec:lgs}

We show how 
the method of proving SN by GS is equally established 
for a labelled system $\labed \RR$.

\subsection{\betaCS}

We need the 
following kind of CS.
Given a CS $(\Sig,\RR)$, the \betaCS~\cite{IDTS00}
is a CS $(\Sig_\beta,\RR_\beta)$  
where 
\begin{itemize}
\item $\Sig_\beta$ is an extension of \Sig extended by
the function symbols for all $\vec a, b \in \Ty_0$
\[
@_{\vec a,b} : (\vec a \to b), \vec a \to b
\]
\item $\RR_\beta$ is obtained from \RR by 
adding
the \W{\beta-rules}
\[
@_{\vec a,b}(\vec{x^{a}}.M[\vec x],\; \vec N) \TO M[\vec N]
\]
for all $\vec a, b \in \Ty_0$.
\end{itemize}

The \betaCS preserves and reflects SN of the original CS by GS.
Hereafter, we will omit the subscripts of @.

\Lem
A CS $(\Sig,\RR)$ satisfies GS if and only if the \betaCS $(\Sig_\beta,\RR_\beta)$
satisfies GS.
\oLem
\begin{proof}
For both sides, take the same order $\le_{\Sig_\beta}$ 
which is an extension of $\le_\Sig$,
where $@$ is smaller than any \Sig-symbol,
and apply GS.
Note that $M,N$ are accessible in lhs of the \beta-rule.
\end{proof}

Therefore, to apply GS to $\labed\RR = \labed\Ax\uni\labed\Bx$,
without loss of generality, 
we assume that $(\SiglB\uni\OH,\labed\Bx)$ is a \betaCS. Namely,
$\labed\Bx$ includes the \beta-rules, and \SiglB includes $@$.
Every \SiglB-symbol is not labelled.

\subsection{Outline}

A key idea of GS is to the use of well-known notion  of Tait's 
\W{computability} to show SN.
Computability implies SN.
The computable closure $\CCl_f(\vec t)$ (Def. \ref{def:ccl})
is a set of meta-terms,
which approximate possible reducts of the lhs $f(\vec t)$ of a rule.
Thus, if all term instances of meta-terms in $\CCl_f(\vec t)$ are computable,
we can conclude the target rewrite system is SN.

Following this idea, we prove SN of $\labed\RR$ with $\tolR$ by GS.
Our proof for the labelled case is actually 
the same as the original one~\cite{IDTS00}, 
except for the use of substitutions, i.e.,
to use 
labelled substitutions $\sap t \rho$ for variables and
labelled substitutions $\lsubst t \th$ for metavariables.
instead of normal substitutions.
The difference between the normal computation $\toR$ and 
the labelled computation $\tolR$ is the use of
$\lext\th(t)$ for meta-applications, which is defined in Def. \ref{def:lmsubst}
as
\[
    \lext\th (\va m[\vec t]) )   =
    \sAP{\vec{x}}{u}{\; \lext\th(t_1),\ooo, \lext\th(t_m)}
\quad\text{for }\th:M \mapsto \vec{x}.u.
\]

The original proof~\cite{IDTS00} actually works
in the case of the labelled case $\labed\RR$, by just 
replacing normal substitutions with labelled substitutions.
A main reason of it is that
labelled meta-terms with labelled substitutions forms 
a \Hi{\Sig-monoid} as shown in~\cite[Appendix A]{HSL}, which is an abstract algebraic model
of syntax with variable binding and substitutions~\cite{FPT,free}.
Because of this well-behaved algebraic structure,
basic properties of substitutions are automatically ensured even in 
the labelled case, including
the substitution lemma, interaction between substitution for variables and
that for metavariables.

To make explicit how the proof works for the labelled case, in the following,
we exhibit crucial parts of the proof, i.e.

\begin{enumerate}[noitemsep]
\item The definition of computability.
\item Key properties of computability.
\item A key lemma for computability closure correctness 
in~\cite[Lemma 13]{IDTS00}.
\end{enumerate}

\subsection{Computability}

\newcommand{\SNlab}{\mathbf{SN}(\TO_{\labed\RR})}

Let $\SNlab$ be the set of all strongly normalising terms by
the computation system $\labed\RR$ with $\tolR$.
We use the following definition of computability.
For a type $\tau\in\Ty$,
we now define $\den{\tau}$ as follows:
\begin{meqa}
\den{b} &= \set{ t \in\Tlab_{b} \| 
t \in \SNlab}   \quad\text{for }b\in\Ty_0\\
%\text{ $t$ is SN by}\tolR  } \quad\text{for }b\in\Ty_0\\
\den{a_1,\ooo,a_n} &= \den{a_1}\X\ccc\X\den{a_n}\\
\den{\vec a \to b} &= 
\set{ \vec x.t \in \Tlab_{\vec a \to b} \| 
\text{for all }\vec s \in \den{\vec a},\; 
\hma{\AP{\vec x}{t}{\vec s} } \in  \den{b}}
\end{meqa}

A term $t$ of type \tau is \W{computable} if $t\in\den{\tau}$.

The above definition is standard~\cite{IDTS00,IDTS}, except for that
SN is considered with respect to $\tolR$, 
which uses labelled substitution, and
$@$-terms are actually evaluated through labelled substitution.
Nevertheless, the surface of definition is the same,
which is a reason why our proof is essentially the same as
the original in~\cite{IDTS00}.

\subsection{Key properties of computability}
The following are crucial properties, which is proved 
by the same way as in~\cite{IDTS00}.

\begin{enumerate}\item Every computable term is SN by $\tolR$ (Lemma \ref{th:CompToSN}).
\item Every one-step reduct of a computable term is computable
(Lemma \ref{th:CompPro}).
\end{enumerate}

\Lem[th:CompToSN]
Every $t \in \den{\tau}$ is SN by $\tolR$.
\oLem
\begin{proof}
By induction on $\den{\tau}$.
  \begin{enumerate}
  \item Case $t \in \den{b}$ for $b\in\Ty_0$. $t \in \SNlab$ by definition of computability.
  \item Case $\vec{x^a}.t\in\den{\vec a\to b}$. 
Since variables $\vec {x^a}$ are SN, $\vec {x^a} \in\den{\vec a}$. 
By definition of $\den{\vec a\to b}$,
$
\AP{\vec x}t{\vec x} \in \den b.$
Hence $\AP{\vec x}t{\vec x} \in \SNlab$, which follows
$\vec x.t \in \SNlab$. \qedhere
  \end{enumerate}
\end{proof}

\Lem[th:CompPro]
Let  $\den{\tau}\ni u$. If $u \tolR t$, then $t\in\den{\tau}$.
\oLem
\begin{proof}
By induction on the type \tau.
  \begin{enumerate}
  \item Case $\den{b} \ni u \tolR t$. $t \in \SNlab$ since $u \in \SNlab$ by definition.
  \item Case $\den{\vec a\to b} \ni \vec {x^a}.u$. 
Suppose $u \tolR t$. Take $\vec s\in\den{\vec a}$.
By definition, $\den{ b}\ni \AP{\vec x}u{\vec s}$.
Applying I.H., we have
$$\den{ b}\ni 
\AP{\vec x}u{\vec s} \tolR \AP{\vec x}t{\vec s} \in \den{b}. $$
By definition of $\den{\vec a\to b}$,
we have $\vec x.t\in\den{\vec a\to b}$. \qedhere
  \end{enumerate}
\end{proof}

\subsection{Key lemma}

We say that a substitution $\rho : X \to \Tlab$ for variables is computable if 
for every $x\in X,\; \rho(x)$ is computable.
We say a substitution $\th : Z \to \Tlab$ for metavariables is computable if 
for every $\th : M \mapsto x.u$, $x.u$ is computable.

Let $t  \in \MlabZ$.
For a substitution $\rho : \Var(t)\to\Tlab$ for variables and 
a substitution \th for metavariables, where \th's codomain does not involve
variables in $\Var(t)$ as free,
$\sap{\lsubst t \th}\rho = {\lsubst {t\rho} \th}$ holds, which is proved by 
easy induction of meta-terms.

To prove a key lemma~\cite[Lemma 13]{IDTS00}
for computability closure correctness,
an order $\gee$ is used to comparing function terms defined for 
\cite[Lemma 13]{IDTS00}.
Since the proof for this case is
the same as the original proof,
we omit the proof and the definition of $\gee$, see~\cite{IDTS00} for details.

\Lem[th:ckey]
Assume that $f(\vec l)\in\MlabZ$ is a second-order pattern,
a computable substitution \th for metavariables
and $\lsubst{\vec l}\th$ is computable.
Assume also that if 
$(f, \lsubst{\vec l}\th) \gee (h, \vec w)$ holds
for $h\in\labed\Sig$ and computable $\vec w$, then
$h(\vec w)$ is computable.
If  $\CCl_f(\vec l) \ni r$, then
${\lsubst {r\rho} \th}$ is computable
for every computable substitution $\rho : \Var(r) \to \Tlab$,
where \th's codomain does not involve
variables in $\Var(r)$ as free.
\oLem
\begin{proof}
First we note that without loss of generality, we assume that bound variables in binders
are always fresh.
The condition ``\th's codomain does not involve
variables in $\Var(t)$ as free'' is satisfied in this case.
We prove by induction on the definition of $\CCl_f(\vec l)$.
\begin{enumerate}
\item Case $r= x$ of type $b$ is SN. 
Then ${\lsubst {\rho(x)} \th}=\rho(x)$ is computable by assumption.
\item Case $r= \vec{x^a}.t$ of type $\vec a\to b$. 
Now $\sap{\lsubst r \th}\rho = {\lsubst {r\rho} \th}$  holds.
Take $\vec s\in\den{\vec a}$. Then
\[
\AP{\vec{x}}{\lsubst{t\rho}\th}{\vec s} \tolR
\sAP{\vec{x}}{\lsubst{t\rho}\th}{\vec s}
= \sap{\lsubst{t}\th}{(\sub{\vec x}{\vec s}\uni\rho)}
\;\in\; \den b \subseteq \SNlab
\]
where ``$\in$'' is by I.H.
Again by I.H., $\lsubst{t\rho}\th$ is computable, 
hence in $\SNlab$. 
$\vec s$ are also computable, hence in $\SNlab$. 
Thus $\AP{\vec{x}}{\lsubst{t\rho}\th}{\vec s}$ is in $\SNlab$, so
$\AP{\vec{x}}{\lsubst{t\rho}\th}{\vec s} \in \den b$.
Therefore,  $\vec{x}.\lsubst{t\rho} \th = \lsubst{\vec{x}.t\rho} \th
\in \den{\vec a\to b}$.

\item Case $r = M[\vec s]$ with $M : \vec a \to  b$.
Now $\sap{\lsubst r \th}\rho = {\lsubst {r\rho} \th}$  holds.
We assume $\th : M \mapsto \vec x.w$.
Then
\[
\sap{\lsubst{M[\vec {s}]}\th} \rho =
\lsubst{M[\vec {s\rho}]}\th = 
    \sAP{\vec x}w{\lsubst{\vec {s\rho}} \th} 
\]
By I.H., $\lsubst {\vec {s\rho}} \th$ is computable.
By assumption, $\vec x.w$ is computable. So
$$\den b \ni
\AP{\vec x}w{\lsubst{\vec {s\rho}} \th} \tolR \sAP{\vec x}w{\lsubst{\vec {s\rho}} \th}$$ 
which is in $\den b$ by Lemma \ref{th:CompPro}.

\item Case $r=h(\vec w)$. Same as in~\cite[Lemma 13]{IDTS00}. \qedhere
\end{enumerate}

\end{proof}

\subsection{SN by GS}
Taking \rho to be the identity substitution in Lemma \ref{th:ckey}
under the same assumptions, we have that 
if  $\CCl_f(\vec l) \ni r$, then
${\lsubst {r} \th}$ is computable.
Then the main result follows from it and~\cite[Lemma 14]{IDTS00}.

\repeatProp[th:lGS]{\lGS}

 \section{A Higher-Order Semantic Labelling}\label{sec:sl}

We prove the postponed 
proposition.

\repeatProp[th:pred-lab]{\PredLab}
\medskip

We prove that every 
infinite \RR-reduction sequence
is strictly simulated by a labelled reduction sequence.
We actually prove
an infinite reduction-plus-subterm-step sequence on minimal non-terminating terms
is strictly simulated by a labelled reduction sequence
because without loss of generality, we can assume that the infinite \RR-reduction sequence
is a minimal one.
A minimal non-terminating term is a non-terminating term
whose strict subterms are SN~\cite{Gramlich}. Let \SN be the set of all strongly normalising terms by
the computation system \RR.
We define the \presheaf of minimal non-terminating terms by
$
\Tinf \deq 
\set{ t \in \TsigV \| 
t \not\in \SN  \;\;\&\;\;
  (\A s \in \TsigV.\; t \;\tgtS\; s \text{ implies }s \in \SN) }.
$
We denote by $\tooRat{\gt\!\root}$ a many-step computation step below the root position,
by $\toRat{\root}$ a root computation step (i.e., rewriting the root position).
The following is well-known.

\Lemma[th:min-seq]{\Minseq}{
  For every $u \in \Tinf$, there exist
\begin{itemize}\item a rule $Z \tpr  l \TO r : b \in \RR$,
        \item an assignment $\th: Z \to \SN$,
        \item a position $p$ of $\subst r\th$,
  \end{itemize}
  such that 
$
u \;\tooRat{\gt\!\root}\; \subst l\th \;\toRat{\root}\;
  \subst r\th \;\tgeS\; 
  \subterm{(\subst r \th)}p \;\in \Tinf,
$
{
and $\subst l\th\in \Tinf$.}
  \oLemma}
\Proof
  Let $u \in \Tinf$ and $Q$ an infinite reduction sequence
  \[
  u = u_0 \toR u_1 \toR \ccc
  \]
  Since all strict subterms of $u$ are SN, $Q$ must contain a root
  computation step.
  Suppose the first root computation step appears as (1) in
$
 u \;\tooRat{(2)}\; \subst l \th \;\toRat{(1)}\; \subst r \th
$
  where $f(\vec{x_1}.l_1,\ooo,\vec{x_n}.l_n) = l$.
  Since (2) is below the root step computation, it must be
  \[
  u = f(\vec{x_1}.t_1,\ooo,\vec{x_n}.t_n) \tooR
  \subst{f(\vec{x_1}.l_1,\ooo,\vec{x_n}.l_n)}\th = \subst l \th
  \]
  and for each $t_i$ $(i=1,\ooo,n)$,
  \[
  \SN \ni\;\; t_i \tooR \subst{l_i}\th \;\;\in \SN
\]
  by the assumption that all strict subterms of $u$ are SN.
  Since $l_i$ is a second-order pattern,
  we see $\th(M) \in \SN$ for every metavariable $M$ in $l$.
  Since $\subst r\th$ is not SN,
there exists a position $p$ such that
  $
  \subst r\th \;\tge\; \subterm{(\subst r\th)}p \;\in \Tinf.
  $
  \QED

\Lemma[flem:thirteen]{\Thirteen}{
  Let 
$\th : Z \rTo \TsigV,\; t \in \MZ$,
and {$t$ is a second-order pattern}.
If $\Fun(t) \subseteq \SigA\uni\OH$, 
      $\trace(\subst t\th) = \aext{(\phith)}(t)$.
      \label{flem:thirteen-eq}
\oLemma}
\begin{proof}
We prove by induction on the structure of the meta-term $t$.

\begin{enumerate}[leftmargin=*]
\item If $t=x$, both sides are $x$.
\item Case $t = M[{\vec x}]$, 
where $\vec x$ are distinct bound variables
because $t$ is a second-order pattern.
Suppose $\th : M \mapsto \vec y.u$. 
\begin{meqa}
 lhs &= \trace (\ext\th( M[{\vec{x}}]  ))
= \trace (u \set{\vec y \mapsto \vec x})
\\
rhs &= \aext{(\phith)}(M[{\vec x}]) 
= \trace(u) \set{\vec y \mapsto \vec x}
= \trace (u \set{\vec y \mapsto \vec x})
= lhs
   \end{meqa}

\item Case $t=f(x.s)$. 
The general case $t=f(\vec x_1.s_1,\ooo,\vec x_n.s_n)$ is proved 
similarly, hence omitted. 
Since
$f\in\SigA\uni\OH$, 
we have by I.H.
\[
lhs = 
f( x.{\trace (\subst s\th)} ) 
= f(x.{\aext{(\phith)}(s)} )
= rhs
\qedhere \]
\end{enumerate}
\end{proof}

The assumption that $t$ is a second-order pattern is essential in the above
lemma. Without it, the case (ii) does not hold because
\tra and substitution for metavariables do not commute for non-patterns.

The labelling $\labtr$ commutes with substitution for metavariables 
in the following sense.

\Lemma[flem:key]{\fKey}{
Let
$\th : Z \to \SN$ be an assignment.
If a meta-term $r$ satisfies the \SigA-layer condition,
then
$
\labtr({\subst r \th}) \;=\;
\lsubst{\lab{(\phith)}(r)}{\abr}.
$
\label{flem:sixteen}
\oLemma}
\Proof
By induction on the structure of the meta-term $r$.

\begin{enumerate}[leftmargin=*]
\item Case $r=x$, both sides are $x$.
\item Case $r= M[\vec s]$. Suppose $\th:M\mapsto \vec y.u$.
\begin{meqa}
lhs 
&= \labtr ({\subst {M[{\vec s}]} \th})
= \labtr (u{\sub{\vec y}{\vec{\subst s\th}}})  
\\
rhs &=
\lsubst{\lab{(\phith)}(M[{\vec s}])}{\abr}\\
&= \lsubst{M[\vec {\lab{(\phith)}\; (s)}]} {\abr}
= \labtr\smultX{\vec y}{u}{\, \vec {\lext{\abr} \lab{(\phith)}\;(s)}}
\\
&= \hma{\labtr\smultX{\vec y}{u}{\, \vec {\lext{\abr} \lab{(\phith)}\;(s)}}}\\
&= \labtr(\smult{\vec y}{u}{\, \vec {\subst s\th} }) = lhs
\end{meqa}

\item Case $r=f(x.s)$. Case $f(\vec{\vec x.s})$ is similar.
\begin{enumerate}[leftmargin=0em]

\item Case 
$f\not\in\SigA$.
    \begin{meqa}
lhs &= \labtr({f(x.\subst s\th)} )
= f(x.\labtr (\subst s\th))\\
&= f(x.\lsubst{\lab{(\phith)}(s)}{\abr})
\quad \text{by I.H.}\\
&= \lsubst{\lab{(\phith)}(f(x.s))}{\abr} = rhs
    \end{meqa}

\item Case $f\in\Sig_\RRR$. 
Since $r$ satisfies the \SigA-layer condition,
{$\Fun(s)\subseteq\SigA\uni\OH$} and {$s$ is a second-order pattern}.
\begin{meqa}
lhs &= \labtr{(f(x.\subst s\th)) } = f_v(x.\labtr (\subst s\th))
\qquad\text{ where } v = f(x.\trace(\subst s \th) )\\
rhs &= \lsubst{\lab{(\phith)}(f(x.s))}{\abr} \\
&= \lsubst{f_w(x. \lab{(\phith)} (s))}{\abr} 
\quad\text{ where } w = f(x. \aext{(\phith)}{(s)} )\\
&= f_w(x. \lsubst{\lab{(\phith)} (s)}{\abr}   ) 
= f_v(x.\labtr (\subst {s}\th)) = lhs
\end{meqa}
By {Lemma \ref{flem:thirteen}}, 
$\trace{(\subst s \th)} = \aext{(\phith)}(s)$, 
hence $w=v$,
The final decreasing step is by I.H. \qedhere
\end{enumerate}
\end{enumerate}
\end{proof}

The $\trace$ with $\RX$-reduction simulates a \RR-reduction as follows.

\Lemma[fth:plab-mon]{\fPlabMon}{
If $\SN \ni s \toR t$ then
$
\trace(s) \tooRX \trace(t).
$
}
\begin{proof}
By induction on the proof of $\toR$.
\begin{itemize}[leftmargin=*]
\item \textbf{(Rule)} If $\subst l \th \toR \subst r \th$,
for $\th : Z \rTo \TsigV$ is derived from
$Z \tpr l\TO r : b\in \RR$
with $l=f(\vec{\vec x.t})$.
If $\subst l \th, \subst r \th \not\in \SN$, then
it is vacuously true.
Case $f \in \OH$ is impossible. 
\begin{enumerate}[leftmargin=*]
\item Case $f \in \Sig_\RRR$. 
Then $l = f(\vec{\vec x.t}) \TO r \in \RRR$ and $\Fun(l)\union\Fun(r)\subseteq \SigA\uni\OH$.
  \begin{meqa}
lhs = \trace(\subst l \th) &= \aext{(\phith)}(l) 
\qquad\text{by Lemma \ref{flem:thirteen}}\\
&\toRX \aext{(\phith)}(r) 
\qquad\text{since $\aext{(-)}$ is an instantiation of an \Ax-rule}\\
&= \trace(\subst r \th) 
\qquad\text{by Lemma \ref{flem:thirteen}}\\
&= rhs
  \end{meqa}

\item Case $f \in \SigB$. Then $f(\vec{\vec x.t}) \TO r \in \Bx$.
  \begin{meqa}
lhs = \trace(\subst l \th) 
&= \tuple\set{\trace(u) \| \subst l\th \TRACEtoR u \trCond}\\
&= \tuple\set{\trace(u) \| \subst l\th \TRACEtoR u \text{ or }
                           \subst l\th \toR \subst r\th = u  \trCond}\\
&\tooRX \trace(\subst r \th) = rhs
\end{meqa}

\end{enumerate}

\item \textbf{(Fun)}
$f(x.s) \toR f(x.t)$ is derived from $s \toR t$.
(The $n$-ary case is similar.)

\begin{enumerate}[leftmargin=*]
\item Case 
$f\not\in\SigB$. 
Using I.H.,
  \begin{meqa}
lhs = \trace(f(x.s)) = f(x.\trace(s)) \tooRX 
f(x.\trace(t)) = \trace(f(x.t)) = rhs
  \end{meqa}

\item Case $f\in\SigB$.
\begin{meqa}
lhs &= \trace(f(x.s)) = \tuple\set{\trace(u) \| f({{x}.s}) \TRACEtoR u \trCond}\\
&= \tuple\set{\trace(u) \| f({{x}.s}) \TRACEtoR u \;\text{ or }\; 
f({{x}.s}) \toR f({ x. t}) = u \trCond}\\ 
&\tooRX \trace(f(x.t)) = rhs
\end{meqa} \qedhere
\end{enumerate}
\end{itemize}
\end{proof}

Then one-step \RR-reduction can be transformed 
to a many-step trace labeled reduction
possibly using decreasing steps on labels.

\Lemma[fth:sound]{\fSound}{
If $\SN\union\Tinf \ni s \toR t \in \SN\union\Tinf$ then
$$
\labtr(s) \;\TOOlabC\; \labtr(t).
$$
\oLemma}
\Proof
By induction on the proof of $\toR$.
\begin{itemize}[leftmargin=0em]
\item \textbf{(Rule)} Suppose that $\subst l \th \toR \subst r \th$,
for $\th : Z \rTo \TsigV$ is derived from
$Z \tpr l\TO r : b\in \RR$.

If $\subst l \th, \subst r \th \not\in \SN \union \Tinf$, then
it is vacuously true.

Case $\subst l \th, \subst r \th \in \SN \union \Tinf$.
We have a labelled rule 
$\lab{(\phith)}(l) \TO \lab{(\phith)}(r) \in \labed\RR$.
By Lemma \ref{flem:key},
\begin{meqa}
\labtr{(\subst l\th)} =
\lsubst{\lab{(\phith)}(l)}{\abr} \toLabC
\lsubst{\lab{(\phith)}(r)}{\abr}
\;=\; \labtr{(\subst r \th)}.
\end{meqa}

\item \textbf{(Fun)}
$f(x.s) \toR f(x.t)$ is derived from $s \toR t$.

(The $n$-ary case:
$f(\ooo, \vec x.s, \ooo) \toR f(\ooo, x.t, \ooo)$ is derived from $s \toR t$,
is similar).

\begin{enumerate}
\item Case $f\not\in\SigA$. Using I.H., we have
  \begin{meqa}
\labtr(f(x.s)) = f(x.\labtr(s)) 
\cTO^+_{\labed\RR\union\Decl}     
f(x.\labtr(t)) = \labtr(f(x.t))
  \end{meqa}

\item Case $f\in\SigA$.
\begin{meqa}
\labtr(f(x.s)) 
&= \phantom{IMM} f_{f(x.{\trace(s)})}(x.\labtr(s)) \\
&\cTO_{\Decl}  f_{f(x.{\trace(t)})}(x.\labtr(s)) 
\qquad\text{by Lemma \ref{fth:plab-mon}}\\
&\cTO^+_{\labed\RR\union\Decl} f_{f(x.{\trace(t)})}(x.\labtr(t))
\quad\text{by I.H.}\\
& = \labtr(f(x.t)) 
\end{meqa}
\qedhere
\end{enumerate}
\end{itemize}
\end{proof}

This establishes a variation of the higher-order semantic labelling method.

\prProp[th:pred-lab]{\PredLab}
{
Let $t\in\Tinf$. By Lemma \ref{th:min-seq}, there exist
\begin{itemize*}[label={}]
\item a rule $Z \tpr  l \TO r : b \in \RR$,
\item an assignment $\th: Z \to \SN$, and
\item a position $p$ of $\subst r\th$
\end{itemize*}
such that
\[
\Tinf \ni\; t \;\tooRat{\gt\!\root}\; \subst l \th \;\toRat{\root}\; \subst r\th \;\tgeS\; 
\subterm{(\subst r \th)}p \;\in \Tinf
\]
By Lemma \ref{fth:sound},
$
\labtr(t) \;\toolabC\!\!\! \labtr(\subst l \th).
$
By Lemma \ref{flem:key},
we have
$$
\labtr(\subst l\th) = \lsubst {\fnlab_{(\phith)}(l)} {\th}.
$$
Since $\lab{(\phith)}(l) \TO \lab{(\phith)}(r)\in \labed\RR$,
$$
\lsubstn{\lab{(\phith)}(l)}{\th} \toLabC
\lsubstn{\lab{(\phith)}(r)}{\th}.
$$

By Assumption \ref{def:setting}\ref{ass:fA},
$r$ satisfies the \SigA-layer condition.
Therefore, by Lemma \ref{flem:key},
$
\lsubstn{\lab{\phith}(r)} {\th}
 =
\labtr(\subst r \th)
$.
Hence
\begin{meqa}
\labtr(t) \;\TOOlabC\;  \lsubstn{\lab{(\phith)}(r)}{\th} 
= \labtr(\subst r \th)
\;\tgeS&\;
\subterm{\labtr(\substn r \th)}{p} \\
=&\; \labtr(\subterm{(\substn r \th)}{p}).
\end{meqa}
Note that since $\labtr$ just attaches labels, it does not affect the position structure
of $r$.

\medskip
Now we show the contrapositive of the proposition.
Suppose \RR is not SN. Then,
there is an infinite reduction sequence
\[
\Tinf \ni t_1 \;\tooRat{\gt\root} \co \toRat{\root} \co \tgeS\; 
          t_2 \;\tooRat{\gt\root} \co \toRat{\root} \co \tgeS\; \ccc
\]
By the above argument, we have
\[
\labtr(t_1)  \;\TOOlabC\co \tgeS\; 
\labtr(t_2)  \;\TOOlabC\co \tgeS\; \ccc
\]
which is infinite, hence
${\labed\RR\union\Decl}$ is not SN.
}

\end{document}